# A Hybrid Formal Verification System in Coq for Ensuring the Reliability and Security of Ethereum-based Service Smart Contracts

**Zheng Yang [1], Member, IEEE, Hang Lei[1], Weizhong Qian[1]**

[1]School of Information and Software Engineering, University of Electronic Science and Technology of China, No.4 Section 2 North Jianshe Road, Chengdu 610054, P.R. China

Corresponding author: Zheng Yang (e-mail: zyang.uestc@gmail.com).

This research did not receive any specific grant from funding agencies in the public, commercial, or not-for-profit sectors.

**ABSTRACT** This paper reports a formal symbolic process virtual machine (FSPVM) denoted as FSPVM-E for verifying the reliability and security of Ethereum-based services at the source code level of smart contracts. A Coq proof assistant is employed for programming the system and for proving its correctness. The current version of FSPVM-E adopts execution-verification isomorphism, which is an application extension of Curry-Howard isomorphism, as its fundamental theoretical framework to combine symbolic execution and higher-order logic theorem proving. The four primary components of FSPVM-E include a general, extensible, and reusable formal memory framework, an extensible and universal formal intermediate programming language denoted as Lolisa, which is a large subset of the Solidity programming language using generalized algebraic datatypes, the corresponding formally verified interpreter of Lolisa, denoted as FEther, and assistant tools and libraries. The self-correctness of all components is certified in Coq. FSPVM-E supports the ERC20 token standard, and can automatically and symbolically execute Ethereum-based smart contracts, scan their standard vulnerabilities, and verify their reliability and security properties with Hoare-style logic in Coq.

## I. INTRODUCTION

Blockchain [1] is one of the emerging technologies developed to address a wide range of disparate problems, such as those associated with cryptocurrency [2] and distributed storage [3]. Recently, Blockchain-as-a-Service (BaaS), where blockchain services are obtained from third-party providers, has established a strong market presence, and is a promising means of utilizing the blockchain technology. Organizations like Oracle, Microsoft, and IBM have launched BaaS offerings. Many BaaS offerings are based on or have been inspired by the Ethereum platform [4], which is one of the most powerful 2.0 blockchain platforms presently available. For example, Amazon partnered with Kaleido to offer cloud services on which to host an Enterprise Ethereum-based architecture. That makes Kaleido the first managed BaaS offering that is available on Amazon web service (AWS) regions across the world.

While BaaS is certainly a rapidly emerging service, the continued rapid development of BaaS is limited by security issues like those affecting Software-as-a-Service (SaaS) products. Presently, one of the security issues of greatest interest for researchers is the reliability and security of blockchain smart contracts. Here, blockchain smart contracts are script programs that provide a kind of special digital contract where the code is the law, which enables blockchain transactions to be conducted automatically [5]. One of the challenges that must be confronted in the development of secure and reliable smart contracts is that the programming process for these programs differs from that of conventional programs. Here, the source code of smart contracts must represent legal considerations in a manner similar to contracts written in natural languages. Therefore, obligations and terms should be presented in smart contracts explicitly. However, smart contract developers are generally programmers, rather than legal experts, and their grasp of legal obligations and terms is secondary to their grasp of programming. As a result, the programming habits of programmers introduce legal loopholes in smart contracts. These loopholes introduce many classes of subtle bugs in smart contracts, ranging from transaction-ordering dependencies to mishandled exceptions [6], that differ significantly from common bugs because, while common bugs are easily detected owing to faulty program execution, these legal loopholes can result in unintended operations, and can also be maliciously exploited to circumvent obligation limitations without affecting the normal execution of the contract procedure, and ultimately result in direct economic loss to users. For example, some of the largest and best known attacks on smart contracts involved those based on the Ethereum platform, such as the attack on the decentralized autonomous organization (DAO) [7] and the Parity wallet attack [8]. A partial summary of Ethereum-based smart contract attacks are listed in Table 1, which indicates that Ethereum attack cases have already resulted in the loss of billions of dollars worth of virtual currency. Moreover, due to the unique feature of these vulnerabilities, standard software engineering techniques employing such static and dynamic analysis tools as Manticore (https://github.com/trailofbits/manticore) and Mythril (https://mythx.io/) have not yet been proven to be effective at increasing the security and reliability of smart contracts. Thus, an urgent need exists for a verification and validation technology that can guarantee the security and reliability of Ethereum-based services in the most rigorous manner available.

TABLE 1
Partial Summary of Recent Ethereum-Based Smart Contract Attack Cases

| Attack Date | Contract Name | Result |
|---|---|---|
| 06/17/2016 | The DAO | 3.6 million ETH lost [7] |
| 07/19/2017 | Parity | 150 thousand ETH lost [8] |
| 11/06/2017 | Parity | 500 thousand ETH frozen |
| 04/22/2018 | BEC | Crashed to zero value[1] |
| 04/26/2018 | SMT | 16 million SMT lost[2] |
| 05/24/2018 | EDU | Crashed to zero value |
| 05/24/2018 | BAI | Crashed to zero value |
| 06/16/2018 | ICX | All ICX frozen[3] |

A brief comparison of existing verification and validation technologies is presented in the Table 2. We note from this table that formal verification is the only technology that satisfies both reliability and completeness simultaneously. Hence, formal verification is one of the most rigorous theoretical technologies available for ensuring the security and reliability of software systems. Consequently, the Ethereum community has focused on the formal verification of smart contracts, issuing open calls for formal verification proposals [9] as part of what has been described as a "push for formal verification" [10].

---

[1] BEC: https://medium.com/@peckshield/alert-new-batchoverflow-bug-in-multiple-erc20-smart-contracts-cve-2018-10299-511067db6536
[2] SMT: https://medium.com/wolf-crypto/batchoverflow-erc20-vulnerability-5691e42940de
[3] ICX: https://medium.com/@tozex/cryptoasset-industry-problems-smart-contract-security-challenge-cb6aee69d5ef

While formal verification is a rigorous approach, different formal verification technologies have their own peculiar strengths and weaknesses. Among these, higher-order logic theorem proving is one of the most rigorous and flexible technologies for verifying the properties of programs. However, numerous problems regarding reusability, consistency, and automation have been encountered when applying theorem-proving technology to program verification [11]. Moreover, the process of model checking is only applicable to finite state programs. In addition, a number of well-known frameworks and tools, focused on Solidity bytecode on the Ethereum virtual machine (EVM) platform, have been developed recently for the verification of Ethereum smart contracts by symbolic execution. For example, the formal semantic known as KEVM was developed for the formal low-level programming verification of EVM using the K-framework [12]. Since KEVM is executable, it can run the validation test suite provided by the Ethereum foundation. Similarly, a Lem [13] implementation of EVM provides an executable semantics of EVM for the formal verification of smart contracts at the bytecode level. Grishchenko et al.'s [14] and Amani et al.'s [15] applied F* and Isabell/HOL to abstractly formalize the semantics of EVM bytecode. Unfortunately, few of these works formalized the syntax and semantics of Solidity source code, and verified the compilation process from Solidity to respective bytecode. Oyente (https://www.comp.nus.edu.sg/~loiluu/ oyente.html) is an EVM symbolic execution engine written in Python supporting most of the EVM opcodes. Several heuristics-based drivers of the symbolic execution engine are built into the tool for finding common bugs in EVM programs. However, these heuristics may introduce false positives in many cases, and they have not been rigorously proven to accurately capture their bug classes. In addition, it must be noted that most of the many recent tools based on symbolic execution adopt model checking technology as their foundation, and few are developed in a higher-order logic theorem proving system to enable real-world programs to be symbolically executed. In addition, conventional symbolic execution approaches also suffer from limitations such as path explosion, where the number of feasible paths in a program grows exponentially, program-dependent efficiency, and memory, environment, and constraint solving problems [16]. Moreover, nearly all of these methods have high learning thresholds. These problems obstruct formal verification technologies from being applied commercially.

TABLE 2
Comparisons of Different Verification and Validation Technologies

| Technology | Reliability | Completeness |
| --- | --- | --- |
| Traditional code testing | Reliable | Non-complete |
| Symbolic execution | Reliable | Non-complete |
| Traditional static analysis | Unreliable | Non-complete |
| Formal verification | Reliable | Complete |

One of the most important features of the Ethereum platform is that it implements a general-purpose Turing-complete programming language denoted as Solidity that allows for the development of arbitrary applications and scripts that can be executed in the EVM to conduct blockchain transactions automatically. However, most prominent studies have focused on the formal verification of the bytecode of the EVM, and the development of high-level formal specifications for Solidity and relevant formal verification tools has attracted considerably less interest despite its importance in programming and debugging smart contract software. Although some intermediate specification languages between Solidity and EVM bytecode have been developed, such as Scilla [17], Simplicity [18] and Bhargavan et al.'s [19], the Solidity syntax and semantics have not been formalized in a manner that is consistent with official documentation. However, the formal syntax and semantics of programming languages actually play a crucial role in several areas of computer science, particularly in program verification. For advanced programmers and compiler developers, formal semantics provide a more precise alternative to the informal English descriptions that typically pass as language standards. In the context of formal methods, such as static analysis, model checking, and program proving, formal semantics are required to validate the abstract interpretations and program logic (e.g., axiomatic semantics) used to analyze and verify programs. Formal semantics for the involved languages are also a prerequisite for verifying programming tools, such as compilers, type checkers, static analyzers, and program verifiers. In other computer science fields, several studies have focused on developing mechanized formalizations of operational semantics for different high-level programming languages. For example, the Cholera project [20] formalized the static and dynamic semantics of a large subset of the C language using the HOL proof assistant. The CompCert project [21] has conducted influential verification work for C and GCC, and a formal semantics denoted as Clight was developed for a subset of C. This formed the basis for VST [22] and CompCertX [23]. In addition, several interesting formal verification studies have been conducted for operating systems based on the CompCert project. Tews et al. [24] developed denotational semantics for a subset of the C++ language that were presented as shallow embedding in the PVS prover. Igarashi et al. [25] presented a minimal core calculus for Java and Generic Java, and verified the important core properties. A similar study was conducted to prove Java type soundness [26]. In addition, the operational semantics of JavaScript have been investigated [27], which is of particular interest in the present work because Solidity is similar to JavaScript. However, most of these studies focused on specific domains and programming languages, and cannot be readily extended for the verification of blockchain smart contracts.

This paper addresses the above issues by developing a formal symbolic process virtual machine (FSPVM) denoted as FSPVM-E for verifying the reliability and security of Ethereum-based services at the source code level of smart contracts. The work of this paper was primarily inspired by the symbolic process virtual machine KLEE [28], which is a well-known and successful certification tool based on symbolic execution. The proposed FSPVM-E is formulated in Coq because it is one of the most highly regarded and widely employed proof assistants [29]. The present study capitalizes on our progress over the past year for developing a powerful hybrid FSPVM system in Coq to verify smart contract properties on multiple blockchain platforms [11], [30], [31], [32], [33]. The proposed system combines the advantages of virtual machine platforms, static vulnerability scanning, higher-order logic theorem proving, and symbolic execution technology based on an extension of Curry-Howard isomorphism (CHI) [34], denoted as execution-verification isomorphism (EVI) [11], and avoids their respective disadvantages, to symbolically execute real-world programs and automatically verify the smart contracts of Ethereum-based services. The present report systematically illustrates the FSPVM-E architecture, elaborates on the novel features of each component, presents experimental results, and introduces real world application examples. Specifically, the present work makes the following contributions.

- **Formal memory model**: We design and implement a general, extensible, and reusable formal memory framework denoted as the GERM framework to virtualize the real-world memory hardware, basic memory operations, and pointer arithmetic in Coq.

- **Formal specification language for Solidity**: We formalize and mechanize an extensible and universal large subset of the Solidity programming language in Coq with generalized algebraic datatypes (GADTs) [35]. This represents a formal intermediate programming language denoted as Lolisa. The supported subset of Solidity is comparable to the subsets commonly recommended for writing common Ethereum-based smart contracts, and includes the built-in functions of the EVM. It also solves the consistency problem associated with existing higher-order logic theorem proving technologies.

- **Execution and proof engine**: An optimized formal verified interpreter is developed in Coq for Lolisa, denoted as FEther, which connects the GERM framework and Lolisa, and serves as an execution engine to symbolically execute and verify the smart contracts of Ethereum written in Lolisa within Coq with high level evaluation efficiency [33].

- **Assistant tools and libraries**: We provide assistant tools and libraries, including a specification generator, a language translator, a static analysis library, and an automatic tactic library, which are respectively applied to generate dynamic specifications, translate Solidity into Lolisa, store the specifications of standard smart contract vulnerabilities, and provide an automatic evaluation strategy. These tools and libraries significantly improve the degree of automation and the validation efficiency of FSPVM-E.

The above core components of FSPVM-E are applied to virtualize Coq as an extensible and general-purpose toolchain for Ethereum smart contract verification, which reduces the verification workload and learning threshold. In addition, FSPVM-E has the following novel features.

- **Executable and Provable**: Defining FSPVM-E based on the GERM framework allows, theoretically, for formal smart contracts to be symbolically executed, and their properties simultaneously and automatically verified using higher-order logic theorem-proving assistants, which, when conducted in conjunction with a formal interpreter, is comparable to the execution of real-world programs.
- **Extensible and Universal**: Although FSPVM-E is designed specifically for the Ethereum platform, each component includes many general features applicable to other blockchain platforms. Thus, the core functions can be extended to formalize similar programming languages.
- **Trusted and Certified**: The self-correctness of FSPVM-E core components is verified in Coq completely, and the correctness of assistant tools not programmed in Coq is also guaranteed.
- **Hybrid verification system**: In conjunction with Lolisa and the static analysis library, FSPVM-E supports basic static analysis to automatically scan standard vulnerabilities hidden in source code. In addition, FEther supports multiple types of symbolic execution for verification using FSPVM-E. Finally, FSPVM-E provides a debugging mechanism for general programmers to identify the vulnerabilities of source code. In this manner, users of FSPVM-E can flexibly select the most suitable method for analyzing and verifying their programs.

The remainder of this paper is organized as follows. Section II briefly presents the essential concepts of FSPVM-E. Section III describes the overall FSPVM-E architecture. Section IV discusses the methods by which the self-correctness of FSPVM-E is ensured, and also defines the non-aftereffect property to guarantee the correctness of assistant tools that are not programmed in Coq. Section V presents simple case studies to illustrate the advantages and novel features of FSPVM-E for verifying the security and reliability of smart contracts. Section VI discusses the extensibility and universality of FSPVM-E, and provides a preliminary scheme for systematically extending FSPVM-E to support different blockchain platforms. Section VII provides a comparison of FSPVM-E with related work for the formal verification for Ethereum smart contracts. Finally, Section VIII presents conclusions and future directions of research.

## II. FUNDAMENTAL CONCEPTS

Prior to defining the formal specifications of FSPVM-E, it is necessary first to define the fundamental environment FSPVM-E.

TABLE 3
STATE FUNCTIONS OF FSPVM-E

| $M$ | Memory space | $\mathcal{E}$ | Environment information |
|---|---|---|---|
| $\Lambda$ | Memory address set | $\Omega$ | Native value set |
| $\mathcal{V}$ | Verified logic term | $\mathcal{F}$ | Overall formal system |
| $\Gamma$ | Proof context | $\mathcal{C}$ | Code segments |

Table 3 lists the state functions used to calculate commonly needed values from the current state of the program. All of these state functions will be encountered in the following discussion. Components of specific states will be denoted using the appropriate Greek letter subscripted by the state of interest. As shown in Table 3, the proof context [29, 36] (denoted as $\Gamma$, $\Gamma_1$, etc.) refers to the proof environment or local proof theory which contains proved local propositions as additional information, and decorate assumptions by discharge rules that are applied when leaving a context, and the $\Gamma_c$ is adoped as the current logic context. In the following contents, we will use *context* to represent *proof context*. The formal memory space in the context is denoted as $M$, and $\sigma$ represents specific memory states. The context of the execution environment is represented as $\mathcal{E}$, and the *env* and *fenv* are adopted as the current environment that stores the current execution environment information, and the super-environment that which stores the super environment information before the execution environment transformation. The symbol $\rhd$ represents the abstract binary relationship that left logic definition could be unfolded as the right logic expression to get more information in the proof context. This meaning of the symbol will be specified in different specific rules. We assign $\Lambda$ to denote a set of memory addresses, and we adopt $\mathcal{V}$ to represent the verified logic terms in the proof assistant.

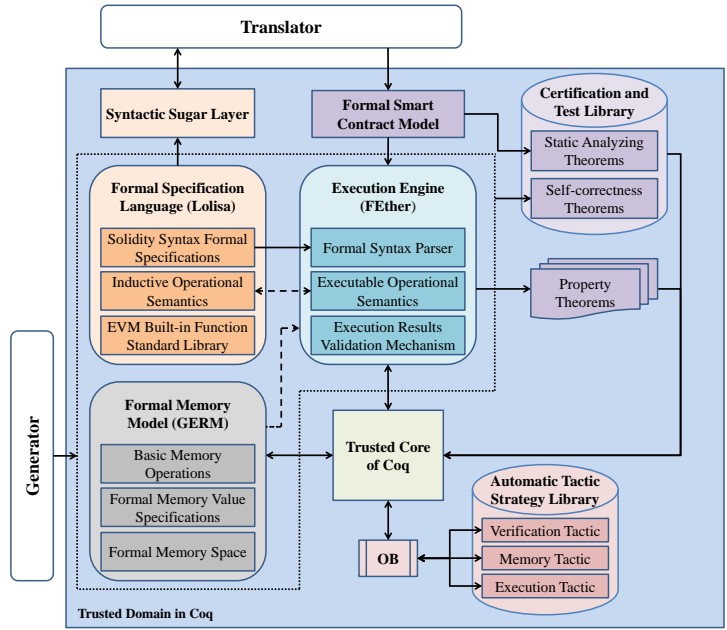

**Fig.1. Architecture of the FSPVM-E verification system.**

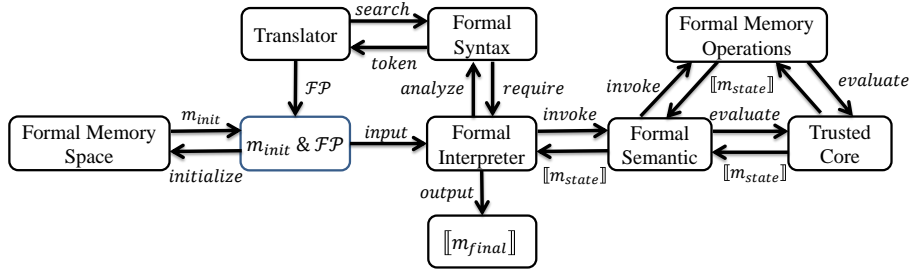

**Fig.2. General workflow of the FSPVM-E.**

Furthermore, we assign $\Omega$ as the native value set of the basic logic system. For brevity in the following discussion, we will assign $\mathcal{F}$ to represent the overall formal system. The operational semantics of Lolisa is abbreviated as $S$, and the evaluation process is denoted as $\Downarrow$. In addition, to avoid ambiguity in the following discussion, we employ the term $Program$ to represent a generic program written in a functional specification language, such as Coq, the term $program_{rw}$ represents a generic program written in a general-purpose programming language $\mathcal{L}$, and $program_{formal}$ represents the formal version of $program_{rw}$ written using the formalized version of $\mathcal{L}$, denoted as $\mathcal{FL}$.

### III. FSPVM-E ARCHITECTURE AND IMPLEMENTATION

The kernel of the FSPVM-E framework was developed entirely in Coq, and Coq was employed as the trusted computation base (TCB) [37]. Here, Coq served not only as a prover for conducting semantic preservation proofs, but also as a programming language for programming all verified components of the FSPVM-E framework. The specification language of Coq includes a small, purely functional language denoted as Gallina, which features recursive functions that operate by pattern matching over inductive types (i.e., ML- or Haskell-style tree-shaped data types). With some ingenuity, Gallina is sufficiently sophisticated for programming a process virtual machine. However, the highly imperative algorithms found in programming language and interpreter textbooks must rewrite in a purely functional style.

The overall architecture of FSPVM-E is illustrated in Fig.1, where a dashed arrow ($\dashrightarrow$) represents a logical dependency relation, and a solid arrow ($\longrightarrow$) represents data transmission. Here, FSPVM-E is designed according to Von Neuman architecture, and it is logically constructed based on two sectors: a foundational sector and an extensible sector. In the foundational sector, the higher-order logic proof environment of Coq is virtualized as the FSPVM-E execution environment, which serves as the logic operating environment, and the trusted core of Coq (TCOC) is virtualized as a central processing unit, and takes Gallina as the logic machine-level language.

The extensible sector consists of four components. The first component is the GERM framework. The second component is Lolisa. The third component is FEther. The final component is the set of assistant tools and libraries, including translator and generator that are developed in untrusted domain using C++. The fundamental component is the formal memory model GERM. It serves as logic memory states to support operational semantics formalization and record all intermediate states during symbolic execution and verification. Based on this formal memory model, the second component, a formal intermediate specification language Lolisa, is able to be defined. It formalizes the selected abstract syntax and corresponding semantics of Solidity language into Coq. Finally, based on the GERM and Lolisa, we can construct the third component that is a formal interpreter FEther. It is used to symbolically execute the formal programs model written in Lolisa.

The general workflow is shown in Fig.2. First of all, FEther will take the initial memory state and formal programs as parameters. Next, the FEther will parse the formal programs according to the syntax of Lolisa. Finally, FEther will invoke the formal operational semantics and interact with GERM to generate a new formal memory state.

These components in the current version of FSPVM-E include over 16,000 lines of Coq source code and 4,000 lines of C++ source code. The specific workload for constructing the FSPVM-E framework is itemized in the Table 4. In this manner, FSPVM-E simulates the process of program execution in the real world, in that a program written in a high-level programming language is interpreted into machine code and executed on the hardware of the operating environment in Coq. The details of these four components are respectively discussed in Subsections III.A to III.D.

TABLE 4
WORKLOAD STATISTICS FOR CONSTRUCTING THE FSPVM-E FRAMEWORK

| Classification | Lines in Coq | Lines in C++ |
|---|---|---|
| Formal memory model (GERM) | 3,849 | 0 |
| Specification language (Lolisa) | 1,004 | 0 |
| Executable interpreter (FEther) | 7,726 | 0 |
| Automatic tactic library | 427 | 0 |
| Self-correctness library | 1,883 | 0 |
| Static analysis library | 1,053 | 0 |
| Generator | 0 | 1,476 |
| Translator | 0 | 2,743 |

## A. GENERAL FORMAL MEMORY MODEL

The GERM framework simulates physical memory hardware structure, including a low-level formal memory space, and provides a set of simple, nonintrusive application programming interfaces using Coq that can support different formal verification specifications simultaneously, particularly at the code level. In theory, it is well suited as a basis for arbitrary high-level specifications in different formal models for program verification.

The overall GERM framework structure is illustrated in Fig.3. The first component of the formal memory model is the formal memory space, which is modeled as a collection of disjoint blocks, and we adopt persistent data structures that support efficient updates without modifying data inplace. Likewise, a monadic programming style enables us to encode exceptions and states in a legible, compositional manner. According to the figure, the GERM framework is based on the TCOC, and can be used as a basis for high-level formal specifications.

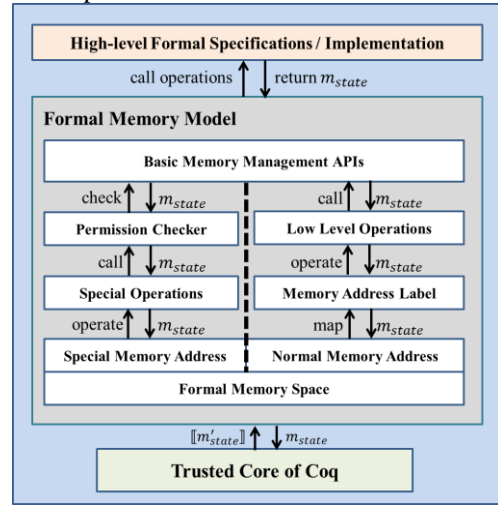

**Fig.3. Architecture of the GERM memory model.**

The data structure of the formal memory space is illustrated by the example given in Fig.4, where each *memory* record field specifies a logic *memory* block with type *value*. Users can define specific memory sizes according to their requirements with the help of assistant tools, which are introduced in Subsection III.D. The details regarding the formal memory space of the GERM framework are presented as follows.

According to the definition of record datatype given by Coq's reference manual [29], each field block $b_{id}$ in Coq has a unique corresponding field name $id$, and the data stored in the $b_{id}$ block of a specific record value $r_v$ can be directly accessed by its corresponding $id$. In *memory*, $b_{id}$ represents the logic memory block, and the corresponding $id$ is formalized as the memory absolute address $M_{address}$. In this manner, the GERM framework statically guarantees that each memory block has a unique $M_{address}$ at the type level. Additionally, in contrast to relevant data structures based on a tree or graph structure [38], which must search all nodes individually to modify a memory block, the GERM framework can access and modify the block directly through its corresponding $M_{address}$.

*Record memory*: Type: = new {
    m_0xinit: value
    m_0x00000000: value
    m_0x00000001: value
    m_0x00000002: value
    m_0x00000003: value
    ...
    m_0x0000000F: value
}

| Memory Address | Data Value |
|---|---|
| m_0xinit | value |
| m_0x00000000 | value |
| m_0x00000001 | value |
| m_0x00000002 | value |
| m_0x00000003 | value |
| ... | ... |
| m_0x0000000F | value |

**Fig.4.** Formal memory space, including the formal specification of memory space in Coq (left), and the real-world physical memory space structure (right).

To improve the flexibility of memory operations, the GERM framework in its current form also provides a label memory address level. In detail, this level first provides a special identifier denoted as a label address $L_{address}$, which is a metavariable defined as an enumeration type in Coq as follows.

$L_{address} ::= \_0x00000000 \mid [\ldots] \mid \_0xFFFFFFFF \mid \Lambda_{special}$

Special address: $\Lambda_{special} ::= \_0xinit \mid \_0xsend \mid \_0xsend\_re \mid \_0xcall \mid \_0xmsg \mid \_0xaddress \mid \_0xblock$

Here, $\Lambda_{special}$ is used as the reference for data structures and functions in the EVM standard library. Accordingly, programmers can define custom mapping strategies $Map_{tac}$ to map $L_{address}$ to $M_{address}$ as follows.

$Map(L_{address}) =$
$match\ Map_{tac}(L_{address})with$
$\mid Error \rightarrow Error$
$\mid OK(M_{address}) \rightarrow M_{address}$
$end.$

Here, if the return value is $Error$, the mapping is deemed to have failed, and the mapping is successful if the return value is $OK(M_{address})$. Similarly, $L_{address}$ can be employed in the GERM framework to accurately model pointer arithmetic. Briefly, users can define pointer objects in high-level specifications, apply $Map(L_{address})$ to reference a location in memory, and obtain the logic expressions stored at that location. The pointer arithmetic mechanism defined in Lolisa based on $L_{address}$ is briefly introduced in Subsection III.B. Moreover, the label address level reserves the extension space for simulating virtual memory and memory isolation mechanisms, such as a memory management unit and a TrustZone.

A benefit of designing FSPVM-E according to von Neumann architecture is that the GERM framework memory space is able to store the program instructions and data values simultaneously and thereby accurately simulate a real-world memory space and guarantee the reliability of the formal memory model. Hence, the memory value type $value$ is inductively defined as rule 1.

$value ::= block_v \rightarrow size \rightarrow env \rightarrow fenv \rightarrow occupy \rightarrow auth \rightarrow value$     (1)

In conjunction with logic memory value $block_v$, as shown in rule 1, the corresponding block size, the current and super logic execution environment information $env$ and $fenv$, respectively, block allocation flag $occupy$, and modification authority $auth$. Currently, $block_v$ supports 14 datatypes, including basic arithmetic data values (*undef*, *integer*, *float*, *Boolean*, *string*, *byte*, *structure*, *array*, and *mapping*), program instructions (*statement*), and pointer objects (*variable pointer*, *parameter pointer*, *function pointer*, and *contract pointer*). One of the benefits of the type constructor of $value$ is that the safety of memory specifications can be readily ensured in the GERM framework. For example, a $block_v$ corresponding to an uninitialized memory block will be initialized as $Undef(tt)$ rather than as a random value. Similarly, all information regarding the current memory state is explicitly stored by the corresponding typing constructors. In this manner, in $\Gamma_c$, the $env$ and $fenv$ record the execution environment information, and the block information indexed by an address can be accessed from current memory state. As abstracted in rule 2, a memory operation $op_{check}$ can modify a specific block by taking current states, address, $env$ and $fenv$ as parameters, and generated a new optional memory state, where $\triangleright$ represents the binary relation that for an arbitrary memory address $addr$ can access a defined logic memory block from the memory state $m_{state}$ of $\Gamma_c$, and $\hookrightarrow$ is represented the evaluation process of $op_{check}$.

$$\frac{\mathcal{E} \vdash env \quad \mathcal{E} \vdash fenv}{\frac{\Gamma_c(m_{state}.addr) \triangleright (block_v, infor_1, \ldots, infor_n)}{op_{check}(m_{state}.addr, env, fenv) \hookrightarrow \llbracket m_v \rrbracket}} \quad (2)$$

As such, conventional memory safety problems, such as buffer overflows and dangling pointers, can be readily exposed in the GERM framework by symbolic execution.

Although the current version of the GERM framework is adjusted to support FSPVM-E, users can extend the $value$ specifications because the GERM framework can function independently with other formal models. Here, as the abstraction of the $value$ definition defined in rule 3 programmers can add new specifications $infor_i$ into the $value$ constructor to represent and store new logic information according to the specific requirements of different projects and situations.

$value: block_v \rightarrow infor_0 \rightarrow infor_1 \rightarrow [\ldots]$
$\rightarrow infor_n \rightarrow value,$     (3)

Finally, we provide 9 classes of basic operations in the GERM framework, which include *map*, *initialize*, *read*, *write*, *address offset*, *search*, *allocate*, and *free*. Any higher-order specifications can take the GERM framework as their fundamental logic state, and apply memory operation APIs to formalize memory-based or memory-related higher level specifications in Coq, such as formal semantics and operating systems.

## B. LOLISA

Following the process illustrated in Fig.5, most of the syntax and respective semantics of Lolisa are formalized strictly following the official Ethereum documentation of Solidity version 0.4. Due to limitation of length, the details of Lolisa's formalization have been presented in our online report (https://arxiv.org/abs/1803.09885).

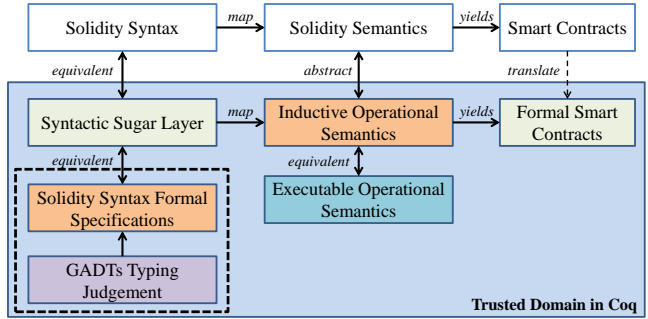

**Fig.5.** Relationships between Solidity and Lolisa.

The specific syntax and semantics of Lolisa are separated as four layers, from bottom to top, including *types*, *values*, *expressions* and *statements*. All reference data structures are formalized in Lolisa because each variable identifier $v_{id}$ is allocated a memory address based on the GERM framework, and can be abstracted as an $\lambda$-application parameter form according to the rule 4 where $x$ is a $\lambda$-bounded parameter and $a$ is a label address.

$$v_{id} := (\lambda (x: L_{address}). x)(a: L_{address}) \quad (4)$$

As such, the syntax of Lolisa not only includes nearly all the characteristic components of Solidity, such as *array*, *mapping*, message, *send*, *transfer*, *address*, *modifier*, *contract*, and *gas* types, but it also contains general-purpose programming language features, such as *multiple return values*, *pointer arithmetic*, *struct*, *field access*, *module*, and the full power of functions, including *recursive functions* and *function pointers*. These abstract syntax components are defined as an inductive type *statement*, and connected by a *list* type. However, there are still some challenges in current version of Lolisa. The first challenge is *inline assembly*. The formalization of *inline assembly* needs accurate formal models of hardware such as registers that has not been supported in current FSPVM-E. Besides, according to our experimental results, the precision of formal floating-point datatype will be lost during unit conversion. Therefore, *inline assembly* and the explicit *ether unit* of Solidity are omitted in Lolisa which are the limitations of current version Lolisa. Currently, if a code segment includes *inline assembly* and explicit *ether unit*, the code needs to be rebuilt manually. We have planned to overcome these two challenges in our future work. In addition, to guarantee structural verification, the *goto* statement, and non-structured forms of switching such as Duff's device [39] are not supported in Lolisa. In this manner, the Solidity can be translated into Lolisa consistently. As explained above, due to limitation of length, the details about formalizations and properties have been presented in our online report [11].

In particular, different from other FSLs of Solidity, such as K framework based Solidity semantics (https://github.com/kframework/solidity-semantics) the formal syntax of Lolisa [11] is defined using generalized algebraic datatypes (GADTs). As indicated by the rule 5, a syntax token $T$ is specified by static type annotations $\tau_i$ for all values and expressions of Lolisa.

$$T :: \tau_0 \to [\dots] \to \tau_n \to Type, \quad (5)$$

This ensures that the formal syntax of values and expressions is more clear and abstract, and facilitates the strict maintenance of type safety for Lolisa expressions. In addition, employing a combination of type annotations facilitates the definition of a very large number of different expressions based on equivalent constructors. The application scope of different statements is limited by different combinations of $\tau_{in}$ and $\tau_{out}$. For example, in the specification of *If* statement defined in rule 6, the input type $\tau_{in}$ of the condition expression has no limitation, but the return type $\tau_{out}$ of the condition expression must be Boolean type $Tbool$. Different from *If* statement, in the specification of *Assign* statement defined in rule 7, the input type $\tau$ and $\tau_0$ of the left and right expressions have no limitation, but the $\tau_{out}$ of the left and right expressions must be equivalent $\tau_1$.

$$If: \forall (\tau: type), expr_{\tau \, Tbool} \to statement \to statement \to statement, \quad (6)$$

$$Assign: \forall (\tau \, \tau_0 \, \tau_1: type), expr_{\tau \, \tau_1} \to expr_{\tau_0 \, \tau_1} \to statement, \quad (7)$$

However, the application of GADTs provides Lolisa with a stronger static type system than Solidity because the type system of Lolisa strictly defines the syntax rules and relevant limitations in the formal specifications as typing judgments. For example, in the *If* statement defined in rule 6, its condition expression is limited by type $expr_{\tau \, Tbool}$, which specifies that the normal form of a specific condition must be a Boolean type. As such, an ill-typed *If* statement, such as

$$\forall s \, s': statement, \left( If \left( Econst \left( Vundef(tt) \right) \right) s \, s' \right),$$

would be disclosed by the type-checking mechanism of Coq, which would then present the error message shown in Fig.6. An additional benefit to the use of GADTs in Lolisa is that the resulting strong static type-checking system assists in discovering errors imported when translating Solidity into Lolisa. This is particularly beneficial because, although syntax errors would be discovered during compilation in the EVM, implementing a compiler mechanism in Coq to check for syntactical correctness when translating Solidity into Lolisa would be an extremely challenging task. Moreover, such errors can seriously affect the evaluation of programs in higher-order theorem-proving assistants.

```
Definition error_type := forall s s', If (Econst(Vundef tt)) s s'.
```

```
Messages  Errors  Jobs

Error:
In environment
s : ?T
s' : ?T0
The term "Econst (Vundef tt)" has type
 "expr (analysis_array_type Tundef) (analysis_array_type Tundef)"
while it is expected to have type "expr (analysis_array_type Tundef) Tbool"
(cannot unify "analysis_array_type Tundef" and "Tbool").
```

**Fig.6.** Simple example of an error message generated by the type-checking mechanism of Coq in response to the ill-typed If statement $\forall\, s\, s':\mathsf{statement},\Big(\mathsf{If}\Big(\mathsf{Econst}\big(\mathsf{Vundef}(\mathsf{tt})\big)\Big)\,s\,s'\Big)$ in Lolisa.

In addition, because Solidity can be translated into Lolisa line-by-line, the strong static type-checking system also assists in discovering ill-typed terms in Solidity source code. To our knowledge, Lolisa is the first formal specification language of Solidity using GADTs.

The semantics of Lolisa are formally defined as big-step operational semantics with two forms: inductive relational forms and executable function forms. The operational semantics of Lolisa defined as the relational form are unified with symbol $S_{rel}$, and the executable operational semantics are represented as symbol $S_{exe}$. $S_{rel}$ is applied to verify properties of Solidity, and $S_{exe}$ is developed as the kernel of symbolic execution engine that can be invoked by FEther. Because Lolisa is employed as a formal intermediate language for Solidity, which should be able to be parsed, executed, and verified in Coq or a similar proof assistant, the semantics of Lolisa are deterministic, and are also based on the GERM framework. The equivalence between the inductive relational forms $S_{rel}$ and the executable function forms $S_{exe}$ of the operational semantics of Lolisa is certified by the following *simulation diagram* theorem, where the *n* represents the step limitation which will be explained in the next section.

**Theorem** *(simulation diagram): Let $\mathcal{E}, M, \mathcal{F} \vdash_{ins} \sigma, opars, env, fenv, b_{infor}$ be the initial evaluation environment, and let $R_{eq}$ represent an equivalence relationship between any two terms. Then, any relational semantic $S_{rel}$ and executable semantic $S_{exe}$ must satisfy the simulation diagram* (Fig.7).

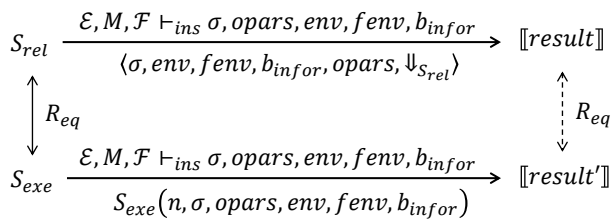

**Fig.7.** Diagram for the correctness certification of Lolisa semantics.

In addition, the development of FEther in Coq and its verification process will be simplified if a $program_{formal}$ written in Lolisa is maintained as a structural program. To ensure this condition, the semantics of Lolisa are made to adhere to the following *program counter* axiom.

**Axiom** *(Program Counter): Suppose that, for all statements s, if s is the next execution statement, it must be the head of the statement sequence in the next execution iteration.*

It must also be noted that the formal syntax of Lolisa contains numerous typing limitations, and is overly complex to accommodate its adoption by general users, as illustrated by the example *If* statement given in Fig.8a and 8b. Therefore, Lolisa is made more intuitive by encapsulating the abstract syntax tree of Lolisa into symbol abbreviations based on the macro-mechanism of Coq, which is denoted as the *notation* mechanism. A *notation* is a symbolic abbreviation denoting some term or term pattern, which can be parsed by Coq automatically. For example, the *notation* shown in Fig.8c can be employed to encapsulate the *If* statement given in Fig.8b, with the result shown in Fig.8d. The use of the *notation* mechanism in Lolisa is illustrated by the components enclosed within the dashed-line box in Fig.5. As a result, the fixed formal syntax components of Lolisa used in verification are transparent to users, and thereby provide users with a simpler syntax. Moreover, this mechanism makes the consistency between real-world languages and Lolisa far more intuitive and user friendly. An additional benefit of this mechanism is that it provides for improved automation of the formalization process. Here, as was conducted when converting Fig.8b to Fig.8d, the syntactic sugar is the interface to connect the translator. The translator is constructed by three modules: lexical analysis module, syntax analysis module and program specification generation module. The lexical analysis module consists of two components: pattern matching sub-module and behavior matching sub-module. The kernel of this module is *Flex* lexical analysis generator that uses regular expression mechanism to form morphemes of characters in source programs. The syntax analysis module is used to generate the abstract syntax tree based on left-child right-sibling binary tree algorithm. Finally, the program specification generation module converts the abstract syntax tree from C++ into the corresponding syntax sugar notations of Lolisa. In addition, these macro-definitions are helpful for promoting extensibility to other programming languages, the details of which are discussed in Section VI.

```
1 function pledge(bytes32 _message) {
2    if (msg.value ==0 || complete || refunded) throw;
3    ...;pledges(numPledges) = Pledge(msg.value, msg.sender, _message);
4    numPledges++;
5 }
```
(a)

```
1 Definition fun_pledge :=
2 (Fun (Some public) (Efun (Some pledge ) Tundef )
3  ( pcons (Epar (Some _message) Tbyte32) pnil))
4  ( nil )
5  (
6    (If ((Ebop forbOfBool
7          (Ebop forbOfBool
8            (Ebop feqbOfInt
9              (Econst (Vfield Tuint (Fstruct _0xmsg msg)
10              (@cons str_name (Nvar values) (@nil str_name) None)))
11              (Econst (Vint (INT I64 Signed 0)))))
12            (Evar (Some complete) Tbool))
13          (Evar (Some refunded) Tbool))
14        )
15      )
16      (Throw;;;) (Snil;;;)
17    );;;
18    ...
19 );;;
20 .
```
(b)

```
1 Notation "s '~>' ss" := (@cons str_name (Nvar s) ss)
2    (at level 59, right associativity).
3 Notation " e1 '(||)' e2 " := (Ebop forbOfBool e1 e2)
4    (at level 58, right associativity).
5 ...
```
(c)

```
1 Definition fun_pledge :=
2 (Fun (Some public) (Efun (Some pledge ) Tundef )
3  ( pcons (Epar (Some _message) Tbyte32) pnil))
4  ( nil )
5  (
6    (If ((SCField Tuint Msg (values ~> \\) None) (==) (SCInt 0)
7        (||) (SVBool complete) (||) (SVBool refunded))
8        (Throw;;;) (Snil;;;)
9    );;;
10   ...
11 );;;
12 .
```
(d)

**Fig.8. Example of macro definitions of the Lolisa formal abstract syntax tree.**

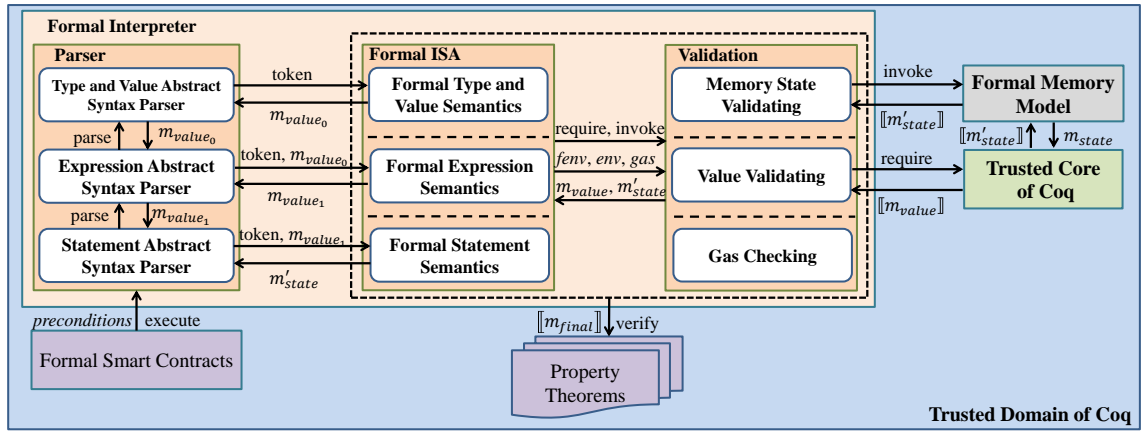

**Fig.9. Architecture of FEther.**

Finally, we have developed a small standard library in Coq that incorporates the built-in data structures and functions of the EVM to facilitate the execution and verification of Solidity programs rewritten in Lolisa using higher-order logic theorem-proving assistants directly. Currently, this standard library is a small subset that includes the syntax components *msg*, *address*, *block*, *send*, *call*, and *requires*. Lolisa is the first mechanized and validated formal syntax and semantics developed for Solidity. The consistency relationship between Solidity and Lolisa is defined according to the following rule 8, where $s$ is the statement of Solidity and $\mathcal{T}$ reprents the translation process from Solidity into Lolisa.

$$\forall\, s, s \in Solidity \wedge \mathcal{T}(s) \in Lolisa \vdash s \equiv \mathcal{T}(s) \qquad (8)$$

### C. FETHER
FEther is an extensible definitional interpreter that is completely developed in Coq based on the GERM framework and Lolisa. The overall FEther structure is illustrated in Fig.9. FEther is entirely constructed in the TCOC, and logically comprises three main components from left to right: a parser, an instruction set architecture (ISA) based on Lolisa semantics, and a validation checking mechanism.

The parser is employed to analyze the syntax of a $program_{formal}$ written in Lolisa, extract the tokens from $program_{formal}$, and invoke corresponding semantics. As mentioned above, Lolisa *types* are used as the type signatures in Lolisa *value* and *expression* layers, and are dependent on corresponding Lolisa *values*. Therefore, the parsing and execution processes of *types* and *values* are combined together.

$$\frac{\begin{array}{ccccc} \mathcal{E} \vdash env, fenv & M \vdash \sigma, b_{infor} & \mathcal{F} \vdash opars & \mathcal{E}, M, \mathcal{F} \vdash P(stt) & \mathcal{E}, M, \mathcal{F} \vdash lib \\ env = set_{gas}\big(init_{env}(P(stt))\big) & fenv = init_{env}(P(stt)) \\ \sigma = init_{mem}(P(stt), lib) \end{array}}{\mathcal{E}, M, \mathcal{F} \vdash FEther\big(\llbracket m'_{state} \rrbracket, env', fenv, args, P(stt)\big) \xRightarrow{execute, T} \langle \sigma', env, fenv \rangle} \quad (9)$$

$$\frac{\begin{array}{ccccc} \mathcal{E} \vdash env, fenv & M \vdash \sigma, b_{infor} & \mathcal{F} \vdash opars & \mathcal{E}, M, \mathcal{F} \vdash P(stt) & \mathcal{E}, M, \mathcal{F} \vdash lib \\ env = set_{gas}\big(init_{env}(P(stt))\big) & fenv = init_{env}(P(stt)) \\ \sigma = init_{mem}(P(stt), lib) \end{array}}{\mathcal{E}, M, \mathcal{F} \vdash FEther\big(\llbracket m'_{state} \rrbracket, env', fenv, args, P(stt)\big) \xRightarrow{execute, \infty} \langle \sigma', env', fenv \rangle \vee env'.(gas) \rightarrow (\neg fenv.(gasLimit)) \xRightarrow{execute, T} \langle \sigma', env', fenv \rangle} \quad (10)$$

Because FEther is a very large function written in the Gallina specification language, FEther differs from the kernel of other real-world virtual machines of high-level programming languages, such as Smalltalk, Java, and Net, which support bytecode as their ISA, and are implemented by translating the bytecode for commonly used code paths into native machine code. In our work, FEther takes the executable semantics of Lolisa as the ISA, which is employed to specify the semantics of the syntax tokens to represent their execution behaviors accurately. Currently, FEther fully supports all semantics of Lolisa.

The validation checking mechanism includes both checking the validation of the results (including memory states and memory values) and checking the execution condition. Because all functions are vulnerable to undefined conditions due to various causes, we develop functions with the help of a monad [40]. Here, all functions are tagged by an *option* type. If a function generates a valid result, the result will be returned in the form of *Some t*. Otherwise, it will be returned as an undefined value *None*. In addition, the symbol $\llbracket t \rrbracket$ represents a term $t$ tagged by the *option* type.

To avoid the execution of infinite loops in programs, FSPVM-E also adopts a *K*-step limitation mechanism, where, similar to Bounded Model Checking (BMC) [41], we limit FEther to executing a $program_{formal}$ $K$ times at most. This gas mechanism design well suits the BMC approach. Therefore, our implementation uses *gas* to limit the execution of Lolisa programs in FEther. To be specific, The semantics governing the execution of a Lolisa program $program_{formal}$ (abbreviated as $P(stt)$) are defined by the rule 9 and 10, where $\infty$ refers to infinite execution and $T$ represents the set of termination conditions for finite execution. These rules represent two conditions of $P(stt)$ execution. Under the first condition governed by the rule 9, $P(stt)$ terminates once a finite number of steps owing to some reasons, such as exception or natural termination. And the $FEther(\llbracket m'_{state} \rrbracket, env', fenv, args, P(stt))$ represents the specific enter point of the whole FEther interpreter. Under the second condition governed by the rule 10, $P(stt)$ cannot terminate via its internal logic and would undergo an infinite number of steps. Therefore, $P(stt)$ is deliberately stopped via the gas limitation checking mechanism that the execution will be terminated after the *gas* balance is reduced to zero. Here, *opars* represents a list of optional arguments. In addition, the initial environment $env$ and super-environment $fenv$ are equivalent, except for their *gas* values, which are initialized by the helper function $init_{env}$, and the initial gas value of $env$ is set by $set_{gas}$. Finally, the initial memory state is set by $init_{mem}$, considering $P(stt)$ and the standard library $lib$ as arguments.

In addition, according to the construction of Lolisa mentioned previously, FEther is also separated as four levels, and it benefits from Lolisa by achieving low coupling between the executable semantics at same layers and at different layers. Specifically, the executable semantics at same levels are independent of each other, and are encapsulated as modules with a set of interfaces. At different layers, higher-layers' semantics can only access lower-layers' semantics via interfaces, and the implementation details of lower-layer semantics are transparent to higher-layer semantics, which is represented by the dashed-line box in Fig.5. The implementation of higher-layer semantics is also not dependent on the details of lower-layer semantics. In this manner, FEther can be extended, as is discussed in Section VI.

### D. ASSISTANT TOOLS AND LIBRARIES

The assistant tools and automation-assisting libraries of FSPVM-E are used to reduce the manual workload and increase the degree of automation when users deploy FSPVM-E, formalize target smart contracts, and verify properties.

***Assistant Tools.*** As discussed previously, FSPVM-E includes customizable components for implementing the specific requirements of users, such as memory space and label addresses, that must be defined dynamically. Taking the memory space given in Fig.4 as an example, the size of the specific formal memory space depends on the size that users require. The enumeration of memory blocks is conducted manually as rule 11.

$$memory \equiv Record\langle m_{addr}{}^*, m_{value}{}^* \rangle \quad (11)$$

Clearly, defining a specific memory space by enumerating blocks manually would be very tedious work. Fortunately, all dynamic specifications like rule 11 above have fixed abstract models. Therefore, these dynamic specifications can be generated recursively by a generator written in a high-level programming language such as Java or C++. In addition, smart contracts written in Solidity can be translated into Lolisa, and vice versa, with a line-by-line correspondence, which are operations that can negatively impact consistency. Similar to the process of converting Fig.8a--d, we develop a translator as a compiler front end to automatically convert Solidity programs into the macro definitions of the Lolisa abstract syntax tree. The abstraction process is given formally by the rule 12.

$$\Gamma \vdash Tools(r) \hookrightarrow specifications \xrightarrow{\text{yields}} v\,files \qquad (12)$$

Here, in the rule 12, assistant tools function Tools employs a specific user requirement $r$ with type $\mathcal{R}$ as a parameter within the proof context $\Gamma$. Assistant tools then generate the respective formal specifications and export them as $.v\,files$ that can be loaded in Coq directly. In this manner, the mechanical processes of FSPVM-E initialization and translation from Solidity into Lolisa can be completed automatically. However, as introduced above, the assistant tools of current framework are developed in C++ without certification. This limitation will be resolved in our next version by developing them in Coq directly.

***Assistant Libraries.*** In standard manual modeling technology, different formal models with significantly different structures and verification processes can be constructed in various programs. Hence, designing a set of tactics that automatically verifies models in different programs is nearly impossible. However, symbolic execution in FSPVM-E corresponds to both function evaluation and program verification, as indicated by the rule 13.

$$\Omega, M, \mathcal{F} \vdash_{ins} P_{exe} \equiv P_{eval} \equiv P_{verify} \qquad (13)$$

Specifically, the verification process $P_{verify}$ of different programs is equivalent with the respective symbolic execution process $P_{exe}$ in FSPVM-E. In other words, the symbolic execution process unifies the verification processes of different programs in higher-order theorem-proving assistants by simplifying the program evaluation process of FEther with different $programs_{formal}$ in FSPVM-E. Because FEther takes the executable semantics of Lolisa as the ISA, FEther execution constitutes a fixed and finite semantics set $\{\mathcal{S}_{exe_0}, \mathcal{S}_{exe_1}, \dots, \mathcal{S}_{exe_n}\}$. Accordingly, we can design a corresponding automatic tactic set $t_i$ for each executable semantic $\mathcal{S}_{exe_i}$, as rule 14.

$$\forall\,i,\ \mathcal{S}_{exe_i} \leftrightarrow t_i \triangleleft T_i\{st_0^i, st_1^i, \dots, st_n^i\} \qquad (14)$$

This can be expanded to accommodate all possible semantic conditions. We exploit this advantage for designing primitive automatic tactics to automatically execute and verify the properties of smart contracts using the *Ltac* mechanism provided by Coq. As shown in Fig.10, the *Ltac*-based tactic modeling is constructed from three components: *memory operating tactics* (MTS), *execution tactics* (ETS), and *verification tactics* (VTS). As the component names suggest, MTS is applied to evaluate the requests of GERM APIs, ETS is used to simplify the executable semantics of Lolisa, and VTS can simplify the pre conditions and post conditions and complete the mathematical reasoning during property verification. The workflow of the tactic modeling is also defined in Fig.10. For Coq operating in the proof pattern, the $observe$ (OB) function scans the current context $C$ to obtain the current goal. Each component in sequence attempts to capture the operational characteristic of the current goal and select the matching tactics. The selected tactics are combined into a solution tactic $Ltac_i$ that solves the goal in the TCOC. The new context $C'$ is compared with $C$ in $context_{dec}$. If $C'$ and $C$ are identical, the current tactics cannot solve the goal automatically, and the tactic modeling process is terminated. Otherwise, the tactic modeling continuously attempts to simplify the goal of $C'$.

An example of the *Ltac*-based tactic modeling is given as rule 15.

$$
\begin{aligned}
&Ltac\ unfold_{modify} \coloneqq\\
&match\ goal\ with\\
&\mid [\,[\,\mid - context\,[?Y\,(?X:memory)(?Z:value)]\,] \Rightarrow\\
&unfold\ Y\ in\ *;\ cbn\ in\ *\\
&end.
\end{aligned} \qquad (15)
$$

Here, the *unfold_modify* tactic is a sub-tactic of a memory operation that captures parts of the operational characteristics of a specific $write_{dir}$ function, and evaluates $write_{dir}$ using basic built-in tactics.

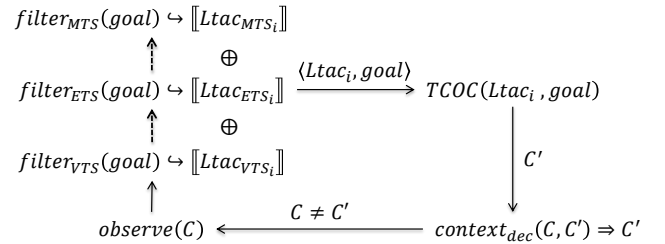

**Fig.10.** Automatic tactic modeling process based on the Ltac mechanism provided by Coq. The process is constructed from three components: memory operating tactics (MTS), execution tactics (ETS), and verification tactics (VTS), where each component in sequence attempts to capture the operational characteristic of the current goal and select the matching tactics that are combined into a solution tactic $Ltac_i$.

Another assistant library is the static analysis library that is used to collect the formal specifications of standard vulnerabilities relevant to coding conventions. This static analysis module is an independent auxiliary module besides the core hybrid verification module. This module is used to check the basic and common vulnerabilities, such as such as integer overflow, unchecked send bug[4] and divide zero, to reduce workload of hybrid verification and improve efficiency of FSPVM-E. These formal specifications are used to construct a static analysis mechanism to quickly scan for bugs in the source code. The details of this library

---

[4] unchecked send bug: http://hackingdistributed.com/2016/06/16/scanning-live-ethereum-contracts-for-bugs/

are discussed in Subsection V.A.

## IV. SELF-CORRECTNESS CERTIFICATION

An essential issue that must be addressed before we can verify smart contracts in FSPVM-E is the self-correctness of FSPVM-E. Because FSPVM-E is employed as the TCB solver for the evaluation and derivation of the formal specification models and properties of smart contracts, the credibility of the analysis results is strongly associated with the correctness of the TCB. Similarly, ensuring the self-correctness of the verification TCB for all static analysis tools is an unavoidable issue. However, the work of Gödel has established that *a logic cannot prove its own consistency*. As such, the fact that verifiers cannot verify themselves presents a paradox. Actually, most theorem proving assistants based on satisfiability modulo theories (SMT) [42] and other types of static analysis tools based on symbolic execution technologies assume that their computation cores are correct, although correctness has not been certified. Therefore, these tools are classified as untrusted in many studies. In contrast, one of the most important features of FSPVM-E is that its self-correctness can be ensured with a very high degree. This is explained in the following subsections. Here, we first demonstrate that the components of FSPVM-E in the trusted domain of Coq can be verified directly. Secondly, we demonstrate that the components in the untrusted domain have no effect on the self-correctness of FSPVM-E from the perspectives of the minimum trusted computational base and the non-aftereffect property.

### A. MINIMUM TRUSTED COMPUTATIONAL BASE

The development of methods to avoid the above-discussed paradox has been conducted since the 1940s, and these approaches have been summarized for proof assistants [37]. Briefly, the manual review of a very small kernel by experts and the *de Bruijn criterion* [43] are two key approaches for ensuring the trustworthiness of a proof assistant. However, the TCB of most static analysis tools is an untrusted black box with a very large kernel that cannot be certified mathematically. Nor does it satisfy the *de Bruijn criterion*. This has been addressed by the Coq team by making the TCOC very small, which has enabled the correctness of the core code to be verified by the manual review of experts. Moreover, the TCOC also completely satisfies the *de Bruijn criterion*. In addition, Harrison [44] has proven the consistency of the HOL Light logical core (or, strictly speaking, a subset of the logical core) using HOL Light itself based on the self-verification concept denoted as reflection, which is also supported in the TCOC. This proof is also completed supported by Coq. Therefore, the TCOC, including the fundamental theory and specific implementation, is currently widely recognized as one of the most reliable trusted cores available.

Besides, another benefit is that the self-correctness of functions and programs developed in Coq using native logic language Gallina, denoted as $\mathcal{F}(Program)$, can be certified in Coq directly.

Taking a very simple example, when we develop an addition *add* in Coq using Gallina language shown as below, we can directly define critical properties in Coq by applying *add* function, such as associative property of addition *plus_assoc* shown below, to certify the self-correctness of this function directly.

*add = fix add (n m : nat) {struct n} : nat :=*

   *match n with*

   *| 0 => m*

   *| S p => S (add p m)*

   *end.*

*Theorem plus_assoc : ∀ n m p : nat, add n (add m p) = add (add n m) p.*

Accordingly, FSPVM-E has a natural advantage compared with other program verification or analysis tools in that its three essential components are wholly developed in Coq, and it employs the TCOC as the TCB. In addition, the core components of FSPVM-E, including the GERM framework, Lolisa, and FEther, are all implemented using the Gallina specification language. Therefore, these are respectively denoted as $\mathcal{F}(GERM)$, $\mathcal{F}(Lolisa)$, and $\mathcal{F}(FEther)$. As such, the correctness of these three components can be directly certified in Coq. Thus, compared with other static analysis tools that must assume the correctness of the TCB, the smart contract verifications that take the proposed FSPVM-E as the TCB must only trust the TCOC. The corresponding verification details are given elsewhere [11, 31, 32]. At present, the core functions of FSPVM-E, which includes 74 theorems and 183 lemmas in Coq, have been completely verified.

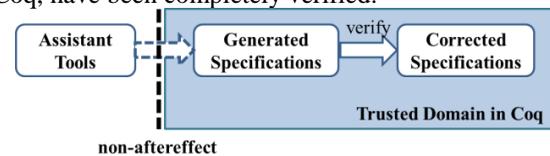

**Fig.11.** Non-aftereffect relation between the dynamically generated specifications of assistant tools with the direct certification of their self-correctness in Coq.

### B. NON-AFTEREFFECT PROPERTY

Although the kernel of FSPVM-E in Coq is verified, the assistant tools are developed in the untrusted domain using general-purpose programming languages. Therefore, these tools are obviously vulnerable to incorrect specifications that can have an impact on the correctness of FSPVM-E. However, the correctness of assistant tools is difficult to audit and certify. Fortunately, the relationship between the assistant tools and their respective results satisfies the non-aftereffect property. Here, although the assistant tools are developed in the untrusted world to automatically generate the dynamic situation-dependent specifications of FSPVM-E, these generated specifications are defined using Gallina, and therefore can be verified in Coq straightly. Hence, as illustrated in Fig.11, the self-correctness of generated specifications can be certified in Coq directly. As such, we can conclude that the component of FSPVM-E affected by generated specifications is correct if the generated specifications pass their correctness certifications. As such, the correctness of FSPVM-E is not influenced by the means through which the assistant tools are implemented.

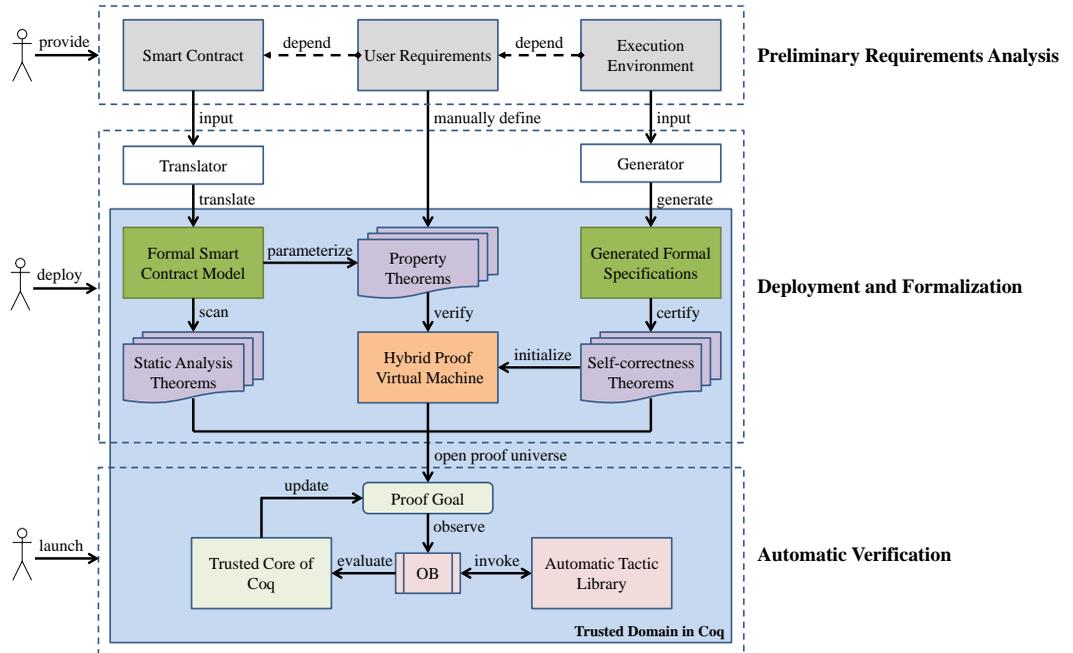

**Fig.12. Summary of FSPVM-E workflow.**

## V. EXPERIMENT

We demonstrate the workflow and novel features of our new toolchain in real-world practice by applying FSPVM-E to formalize and verify a Smart Sponsor Contract (SSC). An SSC is a simple but classic Ethereum sponsor contract demo for new users on the IBM Cloud Laboratory website (https://developer.ibm.com/clouddataservices/2016/05/19/block-chain-technology-smart-contracts-and-ethereum/). Its simplicity makes it an ideal example in the present work. This example offers an additional benefit, in that we have verified this smart contract in our past work using a standard theorem proving approach in Coq [45], and the source code can be downloaded from our Gitee site (https://gitee.com/zyangFV/SSC-manual-verification).

Therefore, we can clearly illustrate the advantages of FSPVM-E by comparison with our past work. Only the necessary code segments are presented in the following discussion to enhance the readability; however, the complete code of this example is given in the Appendix, and can also be downloaded from the cited IBM website. The experimental environment employed 5 identical personal computers with equivalent hardware, including 8 GB RAM and a 3.20 GHz CPU, and equivalent software, including Windows 10 and CoqIDE 8.6.

### A. CASE STUDY: HYBRID VERIFICATION SYSTEM
***Workflow.*** The general workflow of FSPVM-E can be defined in conjunction with Fig.12 that is explained as follows. First, users should provide a smart contract. According to the design requirements of smart contracts, users should abstract them as pre and post conditions of Hoare style properties. Next, users deploy this system in a specific circumstance by applying the generator and translator provided by the assistant tools. The generated formal specifications are certified by the self-correctness theorems stored in the assistant libraries before initializing. These specifications are adopted in FSPVM-E if they satisfy the self-correctness theorems; otherwise, the verification process is terminated. The translator automatically reads the *.sol* file, translates the Solidity smart contract into the respective formal form in Lolisa, defines the program model, and allocates logic memory blocks for all variables. Next, the static analysis module will apply the predefined theorems about basic vulnerabilities to scan the formal version source code of target smart contracts. The analysis results will be printed in the proof goals directly. After that, the formal specification of smart contracts will be parameterized into the properties of program requirements which are defined by users manually. Finally, the users need to open the proof universe, and the hybrid verification engine will finish the verification automatically.

**Fig.13.** Abstract syntax of a code segment definition in FSPVM-E, where $\mathcal{C}$ represents the code segment, $\Gamma[\mathcal{C}]$ denotes the current global context of $\mathcal{C}$, $c_i$ represents a single instruction, $d$ represents a child definition, $\bar{d}$ denotes possible sequences $d_0,\ldots,d_n$, and $D$ is a set of definitions.

**Fig.14.** Partial Smart Sponsor Contract (SSC) model.

A section of the SSC generadted by the translator is shown in Fig.13, where $\mathcal{C}$ represents the code segment, $\Gamma[\mathcal{C}]$ denotes the current global context of $\mathcal{C}$, $c_i$ represents a single instruction, $d$ represents a child *definition*, $\bar{d}$ denotes possible sequences $d_0,\ldots,d_n$, and $D$ is a set of *definitions*. It should be noted that the essence of smart contract segment specifications in Coq is a logic term with a specific logic type *list statement* declared by a *definition*, and FEther is driven by the executable semantics of each abstract syntax. The abstract syntax of code segment definition is also summarized in Fig.13. Hence, FEther has no limitations on the size and continuity of an input $program_{formal}$, and even a single statement can be defined in the proof context of FSPVM-E and symbolically executed by FEther. As an example of the translation process, a formal model of a partial SSC (please see the Appendix) is shown in Fig.14. In addition, the formal version of the *pledge* function of the SSC is presented in Fig.15. We note that the formal version of the *pledge* function is very intuitive, and the critical *requirement* is the pledge operation succeeds if the input value is not 0, and the closed flags *complete* and *refunded* are false. Finally, users need only operate Coq in its proof pattern, and launch FSPVM-E to automatically analyze and verify the smart contracts based on the requirements specified by the user. We provide the following three basic analysis and verification mechanisms in FSPVM-E: static analysis, hybrid verification, and a debugging mechanism. In *Static Analysis Mechanism* subsection, we will introduce the process that static analysis module analyzes the *pledge* function and finds out the basic vulnerabilities. In *Hybrid Property Verification with Hoare Style* subsection, we will illustrate the details that hybrid verification engine verifies important properties of SSC smart contract.

***Static Analysis Mechanism.*** The static analysis mechanism is an independent module, which is embedded within FSPVM-E to identify standard vulnerabilities related to coding conventions and coding mechanisms in smart contracts, such as *integer overflow* and *unchecked send bug*, rather than the logic of the service design. Of course, these vulnerabilities can be accurately identified by defining and verifying corresponding theorems, but this process generates a heavy and tedious workload. However, these vulnerabilities have distinct features that clearly distinguish them from logical vulnerabilities, and are accordingly identified easily and quickly using conventional static analysis technology. Therefore, static analysis is a better choice for supporting this type of security. Moreover, the line-by-line translation between Lolisa and Solidity ensures the consistency between smart contracts and corresponding formal models.

As introduced in Subsection III.D, FSPVM-E includes a static analysis library that contains standard vulnerability features and scanning functions. The abstract definition of *scan* function is defined in rule 16. Here, a *scan* function takes a signed *option* type that the all code segments of context $\Gamma$ will be scanned in the static analysis module.

$$\Gamma[\mathcal{C}] \vdash scan(program_{formal}, feature_i) \Rightarrow [\![c]\!] \qquad (16)$$

If the *scan* function locates vulnerabilities, it will return $Some\ c_{error}$. Otherwise, it will return $None$. For example, if we want to check if an *unchecked send bug* problem exists in the *pledge* function, we need only apply the corresponding scan function to validate the source code, as shown in Fig.16. A scanning result of $None$ indicates that the target $program_{formal}$ contains no code segment with a potential risk of the corresponding feature $feature_i$. Otherwise, the code segment with the potential vulnerability will be located in the proof context. We have encapsulated rule 16 in the current version of FSPVM-E with fixed features to obtain the following particular scan functions.

$$scan_{feature_i} :: list\ statement \rightarrow option\ statement$$
$$scan_{feature_i} := \lambda\ (p: list\ statement).\ scan(p, feature_i)$$

In this manner, the generator can automatically replace $p$ with the target $program_{formal}$ in all feature scanning functions, and automatically analyze $program_{formal}$ according to all of its vulnerability features. This mechanism

```
Definition fun_pledge :=
(Fun (Some public) (Efun (Some pledge ) Tundef)
  ( pcons (Epar (Some _message) Tbyte32) pnil))
  ( nil )
  (
    (If ((SCField Tuint Msg (values ~> \\) None) (==) (SCInt 0) (||) (SVBool complete) (||) (SVBool refunded))
        (Throw;;;) (Snil;;;;)
    );;
    ((Econst (@Vmap Iuint (Tstruct Pledge) pledges (Mvar_id Iuint numPledges) None)) ::=
                  (Estruct ((Vfield Tuint Msg (values ~> \\) None) ;>
                            ((Vfield Taddress Msg (sender ~> \\) None) ;>
                            (str_cnil Pledge))))
    );;
    (Fun_call (Efun (Some _0xsend) Tbool) (pccons (SCField Tuint Msg (addr ~> \\) None) pcnil));;
    ((SVInt numPledges) ::= ((SVInt numPledges) (+) (SCInt 1)));;;;
);;;
.
```

**Fig.15.** Formal version of the pledge function of the SSC.

```
Lemma send_check_bug :
  scan_send_check_bug 100 fun_pledge = None.
Proof.
  scan fun_pledge.
Abort.
```

```
1 subgoal
___________________________________(1/1)
Some
  (Fun_call (Efun (Some _0xsend) Tbool)
            (pccons (SCField Tuint Msg (addr ~> \\) None) pcnil)) = None
```

**Fig.16.** Simple case study of scanning send check vulnerability.

```
Lemma pledge_false : forall k k_val m m1 m2 m3 m4 money msgnum blc gs,
  (env : environment) (s : statement) cp rf num,
  let (_, _, cur, dn) := env in
  m = init_msg init_m money msgnum blc gs smartSponsor msg ->
  m1 = write_dir m  (Bool (Some cp) cur 2 public occupy) complete  ->
  m2 = write_dir m1 (Bool (Some rf) cur 2 public occupy) refunded  ->
  m3 = write_dir m2 (Int (Some (INT I64 Unsigned num)) cur 2 public occupy)
numPledges  ->
  m4 = mem_address (mem_msg m3) ->
  (money = 0)%Z \/ cp = true \/ rf = true) ->
  k > 20 ->
  k_val > 20 ->
  test k k_val m4 (Some nil) env env fun_pledge
  = Some init_m'
.
Proof.
  time(
  intros; init_memory_state env; rewrite_mem'; step_multi).
Qed.
```

```
No more subgoals.

Messages ⌄  Errors ⌄  Jobs ⌄

Tactic call ran for 0.861 secs (0.764u,0.s) (success)
```

**Fig.17.** Static symbolic execution process for verifying the pledge function with abstract symbol arguments.

```
Lemma pledge_false' : forall k k_val m m1 m2 m3 m4 msgnum blc gs,
  (env : environment) (s : statement) num,
  let (_, _, cur, dn) := env in
  m = init_msg init_m 0 msgnum blc gs smartSponsor msg ->
  m1 = write_dir m  (Bool (Some true) cur 2 public occupy) complete  ->
  m2 = write_dir m1 (Bool (Some true) cur 2 public occupy) refunded  ->
  m3 = write_dir m2 (Int (Some (INT I64 Unsigned num)) cur 2 public occupy)
numPledges  ->
  m4 = mem_address (mem_msg m3) ->
  k > 20 ->
  k_val > 20 ->
  test k k_val m4 (Some nil) env env fun_pledge
  = Some init_m'
.
Proof.
  time(
  intros; init_memory_state env; rewrite_mem'; step_multi).
Qed.
```

```
No more subgoals.

Messages ⌄  Errors ⌄  Jobs ⌄

Tactic call ran for 0.222 secs (0.187u,0.s) (success)
```

**Fig.18.** Concolic symbolic execution process for verifying the pledge function.

provides FSPVM-E with the following advantages. First, it provides an optional choice for users to check individual security vulnerabilities efficiently. Second, higher-order logic can be employed to formalize specifications that are more complex than standard specifications. Thirdly, the static analysis library can be extended by adding new specifications and functions for new vulnerabilities. Finally, as discussed in Subsection IV.A, this mechanism is completely developed in Coq, so its correctness can be certified in Coq. Currently, the demo version of FSPVM-E can scan integer overflow, stack overflow and unchecked send bug.

Of course, like similar conventional mechanisms, this mechanism also has the potential for issuing false alarms and providing insufficient error reporting. Therefore, it is provided only as an optional and independent assistant mechanism for users to employ prior to property verification. To our best knowledge, this is the first time that a static analysis mechanism has been embedded within a Coq proof system for Ethereum validation.

***Hybrid Property Verification with Hoare Style.*** As discussed in Section I, potential legal loopholes in smart contracts represent subtle bugs that cannot be detected by standard scanning technologies because these loopholes are closely related to the logic of the specific source code. Solving this problem by means of automated theorem proving represents the most important core function of FSPVM-E. Here, FSPVM-E automatically verifies the requirement properties defined by users based on higher-order logic using Hoare style proof derivations according to the rule 17 which contains three parts.

$$P\{m_{init}\}FEther(m_{init}, program_{formal}, *)Q\{m_{final}\} \qquad (17)$$

First, the precondition $P$ is defined by the initial memory state $m_{init}$. The $m_{init}$ stores the essential constraints that are formalized based on smart contracts design requirements.

Second, the $FEther\ (m_{init}, program_{formal}, *)$ is the entry point of the symbolic execution engine. $FEther$ should take $m_{init}$, the target smart contract $program_{formal}$, and other inputs as parameters.

Third, the postcondition $Q$ is defined by the expected final memory state $m_{final}$, which represents the expected obligation of $program_{formal}$.

This method has no strict restrictions on the specification form of initial and final memory state definitions. The property formalization process will be given as a case study in the next subsection.

Because the Hoare logic derivation is equivalent to the trusted execution of operational semantics, the execution of FEther can be seen as a derivation process based on Hoare logic following the executable semantics of Lolisa.

$$P\{m_{init}\}c_0 \xrightarrow{FEther(m_{init}, c_0, *)} \begin{cases} \xrightarrow{FEther(m_{init}, c_0, *)} Q_0^0\{m_0^0\}c_1 \\ \xrightarrow{FEther(m_{init}, c_0, *)} Q_0^1\{m_0^1\}c_1 \rightarrow \\ \quad ... \\ \xrightarrow{FEther(m_{init}, c_0, *)} Q_0^i\{m_0^i\}c_1 \end{cases}$$

$$\begin{cases} c_n Q_n^0\{m_n^0\} \\ c_n Q_n^1\{m_n^1\} \overset{?}{\leftrightarrow} Q\{m_{final}\}, i \geq 0, j \geq 0 \\ \quad ... \\ c_n Q_n^j\{m_n^j\} \end{cases} \qquad (18)$$

The inference process is given by the rule 18. Beginning with $m_{init}$ as the precondition of program verification, FEther generates all possible proof subgoals, and logically modifies the current memory state $m_{i-1}^j, (j \in \mathbb{N})$ according to the semantics of each statement $c_i$ to generate all possible new postconditions $Q_i^j\{m_i^j\}$ (i.e., the preconditions of $c_i$). The theorems need only judge whether the final output memory state $m_n$ obtained after executing the final statement matches the correct memory state $m_{final}$. Specifically, users need only apply the automatic tactics to automatically complete the symbolic execution of $program_{formal}$, and prove the equivalence between the results of $FEther\ (m_{init}, program_{formal}, *)$ and $Q\{m_{final}\}$. If the equivalence is true, $program_{formal}$ satisfies the respective property theorem.

Because the FEther is a black box for users, general programmers need only be familiar with the mechanical process of applying assistant tools and defining the initial and final memory states with Hoare style logic. Therefore, FSPVM-E reduces the difficulty with which general programmers can apply higher-order theorem proving technology to verify their smart contracts in Coq. In addition, users can alter the symbolic execution process (including static, concolic, and selective symbolic execution) during the verification process by defining the preconditions in different ways. This is explained in the following subsections taking the *pledge* function of the SSC as an example.

### 1) Static Symbolic Execution.

The basic verification process is based on conventional symbolic execution. When the initial arguments are inductively defined with quantifiers such as $\forall$ and $\exists$, the verification engine FEther will follow the logic of the source code to traverse all cases that satisfy the preconditions. Specifically, according to the requirement given in **Workflow** subsection, if one of the sponsor termination flags *complete* and *refund* is set as true or the donation amount is zero, the smart contract must be discarded. Following this requirement, the property formalization is given bellow.

First, the Lemma *pledge_false* defined in Fig.17 initializes $\big(\text{Bool}\ (\text{Some} ? X)\big)$, $\big(\text{Bool}\ (\text{Some} ? Y)\big)$, and $(\text{INT I64 Signed}? Z)$ by specifying $? X$, $? Y$, and $? Z$ as inductive values representing all possible conditions of initial termination flags and the donation amount, namely, as $\forall (cp: bool)\big(\text{Bool}\ (\text{Some}\ cp)\big)$, $\forall (rf: bool)\big(\text{Bool}\ (\text{Some}\ rf)\big)$, and $\forall (num: int)(\text{INT I64 Signed}? Z)$, respectively. Specifically, the initial memory states $m$ to $m_4$ that satisfy the *pledge* function and other essential preconditions are defined first.

Second, according to the requirement, the constraint of donation amont, *complete* and *refund* are defined as $\text{money} = 0 \lor complete = true \lor refund = \text{true}$.

Third, the formal specification *fun_pledge* and preconditions are given into the entry pointer of FEther, denoted as *test*. Next, the expected postcondition that the *pledge* function is not applied successfully is defined as *Some init_m'* in Fig.17.

Finally, users need only mechanically apply the compositive automatic tactics to initialize the preconditions as the initial execution environment, symbolically execute *fun_pledge*, and verify the equivalence between the execution result and the postcondition by forward reasoning. This represents the complete property verification process. Programmers can execute and complete this property verification using the automatic tactics of FEther within 0.861 s (right-hand side of Fig.17).

### 2) Concolic Symbolic Execution.

Because Coq is also a kind of functional programming language, it supports the evaluation of specific real inputs. As shown in Fig.18, the entry points *test* and code *pledge* are unmodified, and *complete* and *refunded* are replaced with specific values true and false, respectively. The other constraints are still inductively defined as abstract symbols. The functional correctness with the specific inputs is then proven by the lemma *pledge_false*. Because the inputs are specified, the number of possible execution paths is limited, and the execution time is reduced to 0.222 s (right-hand side of Fig.18). This mechanism leaves extensible space support for standard testing. Users can extend the generator or implement a new test mechanism to automatically generate test script to modify the initial arguments and validate target programs.

### 3) Selective Symbolic Execution.

This verification process allows programmers to extract segments of code from target programs, verify the properties of the selected code segments separately, and apply these verified properties to simplify the verification process of target programs. For example, a code segment $\mathcal{C}$ can be redefined as follows.

$$\mathcal{C} \overset{\text{def}}{=} d_c = [c_0, .., c_i] + \cdots + [c_j, .., c_k] + \cdots + [c_m, .., c_n] + \cdots + [c_y, .., c_z]$$
$$= [c_0, .., c_i] + \cdots + d_i + \cdots + d_j + \cdots +$$
$$+ [c_m, .., c_n] \rhd \Gamma'(d_i \overset{\text{def}}{=} [c_j, .., c_k], d_j \overset{\text{def}}{=} [c_m, .., c_n])$$
$$= [c_0, .., c_i] + \cdots + d_i + \cdots + D_k \rhd \Gamma'\left(d_i \overset{\text{def}}{=} [c_j, .., c_k], D_k \overset{\text{def}}{=} d_j \rhd \Gamma''(d_j \overset{\text{def}}{=} [c_m, .., c_n])\right)$$
$$Verify(\mathcal{C}) = Verify([c_0, .., c_i] + \cdots + d_i + \cdots + D_k)$$

Therefore, the key point is whether $Verify([c_0, .., c_i] + \cdots + d_i + \cdots + D_k)$ is equivalent to $Verify([c_0, .., c_i]) + \cdots + Verify(d_i) + \cdots + Verify(D_k)$.

First of all, any code segment $d$ that satisfies the syntax given in Fig.13 will not be unfolded and executed and verified in FEther according to the rule 19.

$$P\{m_{init}\}c_0 \xrightarrow{FEther(m_{init}, c_0, *)} Q_0\{m_0\}d$$
$$\xrightarrow{FEther(m_0, d, *)} Q_1\{m_1\}c_i \to \cdots \leftrightarrow Q\{m_{final}\} \quad (19)$$

Obviously, we obtain the result $P_d\{m_d\}\ d\ Q_d\{m_{df}\}$, which means that, according to the purposed precondition $P_d\{m_d\}$, $FEther(m_d, d, *)$ will obtain the expected logic state $Q\{m_{df}\}$. In addition, because Coq employs a *call-by-name* evaluation strategy [46], the bodies of all definitions, including functions and values, are stored in their own contexts, and are not evaluated until they are needed in the current proof context $\Gamma_c$. Therefore, any segment $d$ of the logic expression $\varepsilon$ in the current proof context will not be unfolded during the proof process until all instructions prior to $d$ have been executed in the proof context, as indicated by the rule 20, where $\rhd$ denotes the binary

**Fig.19.** Selective symbolic execution process for verifying the pledge function.

relationship that each context has subcontext with enscapluated definitions recursively.

$$\frac{\Gamma \rhd \mathcal{D}\left(d_\Gamma, \Gamma' \rhd \mathcal{D}\left(d_{\Gamma'}, \Gamma'' \rhd \mathcal{D}(d_{\Gamma''}, \dots)\right)\right) \vdash \varepsilon \xrightarrow{reduction} \varepsilon_{nf}}{\Gamma' \rhd \mathcal{D}\left(d_{\Gamma'}, \Gamma'' \rhd \mathcal{D}(d_{\Gamma''}, \dots)\right)) \vdash \varepsilon_{nf} \oplus d_\Gamma} \quad (20)$$

Moreover, if a proposition $Q$ is a subset of a proposition $P$, $P$ can imply $Q$, which is denoted as $\forall\, Q.\, Q \subseteq P \vdash P \to Q$. This is formalized as the rule 21.

$$P\{m_{init}\}\overline{c} \to Q_c\{m_c\}_{\subseteq P_d\{m_d\}}\, d\, Q_d\{m_{df}\}$$
$$\xrightarrow{specify} Q_{d_c}\{m_{df_c}\}\, \overline{c'} \to \cdots \overset{?}{\leftrightarrow} Q\{m_{final}\} \quad (21)$$

Here, if the final logic memory state $Q_c\{m_c\}$ of the instruction set $\overline{c}$ prior to $d$ is a subset of the precondition $P_d\{m_d\}$, which has already been verified, $P_d\{m_d\}$ can be specified as $Q_c\{m_c\}$ by applying $P_d\{m_d\} \to Q_c\{m_c\}$, and $Q_d\{m_{df}\}$ can also be determined according to the specific $P_d\{m_d\}$. This process is executed, and can be automatically completed in Coq using the *eapply*, *eauto*, and *Hint Resolve* tactics. As introduced previously, FSPVM-E takes the GERM logic memory state as the pre and post conditions in the Hoare triple. Therefore, the excepted abstract or specific final memory state of $\overline{c}$ can be taken as the initial memory state of $d$. Obviously, if $\overline{c}$ is correct, its final logic memory state is a subset of its expected logic memory state. The same procedure can be adapted to an arbitrary definition set $D$. Hence, specifying all preconditions of $d$ and $D$ by the input post conditions yields $Verify([c_0, .., c_i] + + \cdots + + d_i + + \cdots + + D_k) = Verify([c_0, .., c_i]) + + \cdots + + Verify(d_i) + + \cdots +$

$+Verify(D_k)$.

As shown by the example in Fig.19, the *pledge* function can be varied by exploiting the selective symbolic execution of FEther. Programmers can extract the core code segment $if\ (msg.value = 0\ ||\ complete || refnd)\ \{throw();\}$ from the *pledge* function and redefine it as a new formal definition $d$ denoted as *fun_pledge_if*, as shown in the red box, which can in turn be independently verified by the *pledge_false_select* lemma. After combining the verified *fun_pledge_if* code segment into the *pledge* function, the verification of the *pledge_false* theorem can be completed by invoking the *pledge_false_select* lemma. Clearly, the *pledge_false_select* lemma can also assist in any other proofs that use the *fun_pledge_if* code segment. Capitalizing on this feature, FSPVM-E is able to improve its reusability and mitigate the effects of path explosion by extracting important or universal code segments, and verifying them separately. The details of this process are discussed in Subsection V.D.

### 4) Debugging Mechanism.
Finally, FEther provides an interactive debugging mechanism for users. Because FEther is developed based on the GERM memory model, the formal intermediate memory states record the proof information, such as logic invariants and expressions, of all variables during the execution and verification process. The proposed mechanism employs the debugging tactic *step_debug* to enable the

**Fig.20. Example** Of the debugging mechanism provided by FSPVM-E.

```
with
| Some true =>
    test k k_val
        {|
        m_init := initData;
        m_send := initData;
        m_send_re := initData;
        m_msg := Str_type _0xmsg
                    (str_mem Taddress (Nvar sender)
                        (str_mem Tint (Nvar values) str_nil)) _0xmsg 3 occupy;
        m_address := Str_type _0xaddress
                        (str_mem Tint (Nvar addr)
                            (str_mem Tint (Nvar balance)
                                (str_mem (Tfid (Some _Send)) (Nvar send)
                                    (str_mem Tint (Nvar gas) str_nil)))) _0xaddres
                            3 occupy;
        m_throw := false;
        m_0x00000000 := initData;
        m_0x00000001 := initData;
        m_0x00000002 := initData;
        m_0x00000003 := initData;
        m_0x00000004 := initData;
        m_0x00000005 := initData;
        m_0x00000006 := initData;
        m_0x00000007 := initData;
        m_0x00000008 := initData;
        m_0x00000009 := initData;
        m_0x0000000a := initData;
        m_0x0000000b := Bool (Some rf) a 2 public occupy;
        m_0x0000000c := Bool (Some cp) a 2 public occupy;
        m_0x0000000e := Int (Some (INT I64 Unsigned num)) a 2 public occupy;
```

```
        m_0x0000005b := initData;
        m_0x0000005c := initData;
        m_0x0000005d := initData;
        m_0x0000005e := initData;
        m_0x0000005f := initData;
        m_0x00000060 := initData;
        m_0x00000061 := initData;
        m_0x00000062 := initData;
        m_0x00000063 := initData |) (Some nil) (Evn 1 None pledge d)
        (Evn l o a d)
        Econst (Vmap pledges (Mvar_id Iuint numPledges) None)
        ::= Estruct
                ((Vfield Tuint Msg (values -> \\) None;>)
                    ((Vfield Taddress Msg (sender -> \\) None;>)
                        (str_cnil Pledge)));;
        SVInt numPledges ::= SVInt numPledges (+) SCInt 1;; ;)
| Some false =>
    test k k_val
        {|
        m_init := initData;
        m_send := initData;
        m_send_re := initData;
        m_msg := Str_type _0xmsg
                    (str_mem Taddress (Nvar sender)
                        (str_mem Tint (Nvar values) str_nil)) _0xmsg 3 occupy;
        m_address := Str_type _0xaddress
                        (str_mem Tint (Nvar addr)
                            (str_mem Tint (Nvar balance)
                                (str_mem (Tfid (Some _Send)) (Nvar send)
                                    (str_mem Tint (Nvar gas) str_nil)))) _0xaddres
                            3 occupy;
        m_throw := false;
```

**Fig.21.** Formal intermediate memory states during the verification process of the pledge function.

step-by-step debugging of a smart contract by manually tracing these intermediate memory states. As shown in Fig.20, the states of all memory blocks at the current break point are printed in the proof context. Programmers can extract a target code segment using selective symbolic execution with debugging tactics, and trace the intermediate memory states to locate bugs. Fig.21 shows the formal intermediate memory states obtained during the execution and verification of the *pledge* function using FEther in the proof context. Then, we can compare the mechanized verification results and the manually obtained results to validate the semantics of Lolisa. In addition, the application of FEther based on Lolisa and the GERM framework also certifies that our proposed FSPVM-E is feasible. Here, the logic memory states during execution can be observed in the proof context.

### B. ADVANCED FSPVM-E FEATURES
The advanced features of the purposed FSPVM-E include *consistency*, *automation*, and *evaluation efficiency*, which are summarized individually in the following subsections.

**Consistency.** The consistency feature of FSPVM-E includes language formalization consistency and execution and verification consistency.

#### 1) Language Formalization Consistency. This consistency feature refers to the line-by-line translation between Solidity and Lolisa following the formalized syntax of Solidity. Many well known higher-order logic theorem proving frameworks, such as deep specifications [23], require researchers to manually abstract or rebuild the resource code of target programs as computational formal specifications in higher-order logic theorem proof assistants, and the overall process depends entirely on the experience, knowledge, and proficiency of researchers rather than on a standardized and mechanized criterion. As a result, the consistency between the formal model and the original

program cannot be ensured formally, which represents one of the most troubling problems associated with higher-order theorem proving technology. As shown in Fig.22, which was extracted from our previous work [33], the abstract formal specification of the *pledge* function in the Gallina specification language is far different from the corresponding source code given in Fig.15 and the Appendix, even though it accurately defines the behavior of the *pledge* function. This flexibility in the abstraction and translation processes leads to a general lack of consistency between the formalization results obtained by different researchers, and the consistency between formal specifications and corresponding source programs is also very difficult to certify. As a result, the formal model runs the risk of misunderstanding the source program logic and implementation, and may import vulnerabilities not existing in the original program, or remove vulnerabilities existing in the source code as an unintended result of the abstraction and translation processes.

```
1  Definition func_pledge (ct : contract) (ms : msg) (mss : message) : contract :=
2      match state_dec (valid_pledge ms ct) return contract with
3      | right _ => set_state throw ct
4      | left _ =>
5          let pled := Infor ms.(value) ms.(sender) mss in
6          let l := build_list ct.(pledges) pled in
7          mkcontract ct.(owner) ct.(benefactor) ct.(refunded)
8              ct.(complete) false false (S ct.(numPledges))
9              l (ct.(balance) + ms.(value)) continue ct.(accounts)
10     end.
```

**Fig.22.** Manually translated version of the formalized pledge function [45].

Although, in current FSPVM-E, the translator is developed in untrusted domain using C++ that we cannot guarantee the correctness of it, as indicated by the following rule

$$program_{rw} \in Solidity \wedge \mathcal{T}(program_{rw}) \in Lolisa \vdash$$
$$program_{rw} \equiv program_{formal} \qquad (22)$$

where the symbol $\mathcal{T}$ represents the translation process, in contrast with conventional approaches, Lolisa has mechanized most syntax and corresponding semantics of Solidity into Coq. In addition, the formal syntax of Lolisa has also been encapsulated into the syntactic abbreviations using the *Notation* macro mechanism, and hides the fixed formal syntactic components. Benefiting from this, theoretically, the formal version of smart contracts is very similar to their original version, so the equivalency of small size smart contracts could be reviewed by experts or third-party plug-ins. Hence, as demonstrated by Figs.14 and 15, smart contracts can be translated from Solidity into Lolisa line-by-line. Therefore, rule 22 above can be transformed into the following rule:

$$program_{formal} \equiv program_{model} \qquad (23)$$

and $program_{formal}$ can be directly executed and verified in Coq with the help of FEther. Accordingly, $program_{formal}$ is actually the corresponding formal specification model $program_{model}$ at the code level. Furthermore, according to logic transitivity, rules 8, 19, and 23 can be combined to obtain the following rule.

$$program_{rw} \in Solidity \wedge \mathcal{T}(program_{rw}) \in Lolisa \vdash$$
$$program_{rw} \equiv program_{formal} \equiv program_{model} \qquad (24)$$

Therefore, the target smart contract $program_{rw}$ is consistent with its formal model $program_{model}$ in the proposed FSPVM-E.

Because a simple translator can translate Solidity into Lolisa line-by-line automatically, and researchers can check the consistency between $program_{formal}$ and $program_{model}$, this process ensures the objectivity of the formal model, and depends in no way on the experience, knowledge, and proficiency of researchers. Thus, this process guarantees consistency between the source program and the respective formal model. Although the equivalence between Solidity and Lolisa cannot be totally guaranteed in current version FSPVM-E due to the limitation of the translator, this is a best effort to preserve the equivalence between two languages. Besides, we also have planned to develop a trusted translator in Coq directly.

*2) Execution and Verification Consistency.* The execution consistency refers to the accuracy with which the symbolic execution of smart contracts in FSPVM-E simulates the actual behaviors of smart contracts when they are executed in the real world. As elaborated by the design and the case studies given above, this feature is successfully facilitated in FSPVM-E by several ways. First, the GERM framework virtualizes the architecture and operations of real world memory hardware, and the semantics of Solidity has been mechanized into Coq as the core of FEther. Moreover, Ethereum smart contracts written in Lolisa can be directly executed in FSPVM-E. Hence, the execution and verification level of FEther functions directly on what is essentially Solidity code rather than byte code, which avoids the risk of errors during compilation. Therefore, the symbolic execution of target smart contracts in FSPVM-E accurately reflects their behaviors in the formal systems of Coq.

*Automation.* As demonstrated by the case studies, another very important feature of FSPVM-E is its high level automation of higher-order theorem proving. The ratio of source code size to proof code size in many conventional higher-order theorem proving approaches (e.g., [23, 47]) varies linearly in the range of about $20:1$ to $40:1$, and the workload can be summarized by the rule 25, including the size of manual formalization and verification.

$$workload_{manual} =$$
$$size(formalization) + size(verification) +$$
$$size(property) \qquad (25)$$

Similar to these conventional approaches, properties must be manually defined in FSPVM-E. However, the size of the source code and the size of the proof code no longer have any direct relationship in FSPVM-E owing to the high level of automation of the formalization and verification processes.

The individual components of the workload for FSPVM-E (i.e., $workload_{FSPVM-E}$) can be analyzed as follows. First, as introduced in the above discussion regarding language formalization consistency, the formalization process is entirely unified as a line-by-line translation from Solidity to Lolisa, and we have already implemented a translator to make the translation process fully automatic. Therefore, $size(formalization)$ is zero. Second, the ability of most mainstream interactive proof assistants, such as Coq and Isabelle, to reduce the workload associated with verification is limited, despite the fact that they provide *tactic* mechanisms to help users design proving tactics to simplify the program evaluation process and construct proofs automatically. This is because the above discussed differences among the different formal models derived from manual design make it difficult to design tactics that can verify formal models in a fully automatic fashion. However, in contrast to conventional verification approaches and frameworks, FSPVM-E standardizes the verification process as the execution of smart contracts in FSPVM-E according to rule 13. In addition, as discussed in Subsection III.D, the finite number of executable semantics and operations in FSPVM-E facilitates the development of design strategy sets to automatically accommodate all possible semantic conditions, and these strategies are encapsulated as three basic tactics. The substantial reduction in $size(verification)$ facilitated by FSPVM-E is well illustrated by comparing the verification of the *pledge_false* theorem shown in Fig.17, which requires only a single line of code, with that obtained by directly verifying *pledge_false* using the conventional tactics of Coq, as shown in Fig.23, which requires 20 lines of complex tactics. dThese three basic tactics can be applied manually or invoked automatically with the help of script programs. According to these features of FSPVM-E, $size(verification)$ can be reduced to zero theoretically.

```
Lemma pledge_false : forall k k_val m m1 m2 m3 m4 money msgnum blc gs
  (env : environment) (s : statement) cp rf num,
  let (_, _, cur, dn) := env in
  m = init_msg init_m money msgnum blc gs smartSponsor msg ->
  m1 = write_dir m (Bool (Some cp) cur 2 public occupy) complete  ->
  m2 = write_dir m1 (Bool (Some rf) cur 2 public occupy) refunded  ->
  m3 = write_dir m2 (Int (Some (INT I64 Unsigned num)) cur 2 public occupy) numPledges   ->
  m4 = mem_address (mem_msg m3) ->
  ((money = 0)%Z \/ cp = true \/ rf = true) ->
  x > 20 ->
  k_val > 20 ->
  test k k_val m4 (Some nil) env env fun_pledge
  = Some init_m'
.
Proof.
  destruct env. intros.
  destruct k; try omega; simpl.
  rewrite H3, H2, H1, H0, H.
  simpl.
  destruct k; try omega; simpl.
  destruct (address_dec _0xmsg _0xmsg) eqn: Haddr_dec; try congruence.
  destruct (str_name_dec (Nvar sender) (Nvar values)) eqn: Heq_sender_values; try discriminate.
  destruct (str_name_dec (Nvar values) (Nvar values)) eqn: Heq_values_values; try congruence.
  destruct H4.
  - (* money = 0 *)
    rewrite H4. simpl.
    destruct cp eqn: Hcp, rf eqn: Hrf; simpl; destruct k; try omega; simpl; eauto.
  - destruct H4; rewrite H4; simpl.
    + (* cp = true *)
      unfold eqInt; rewrite par_valid;
      destruct money eqn: Hmoney; simpl; destruct k; try omega; simpl; eauto.
    + (* rf = true *)
      unfold eqInt; rewrite par_valid; destruct money eqn: Hmoney; simpl;
      destruct cp eqn: Hcp; simpl; destruct k; try omega; simpl; eauto.
Qed.
```

No more subgoals.

**Fig.23. Verification of the pledge function using the built-in tactics of Coq.**

Finally, the above discussion indicates that $workload_{FSPVM-E}$ is related only to the size of the verification properties, i.e., $workload_{FSPVM-E} = size(property)$, which depends on the complexity of the requirements of target programs rather than on the size of their source code.

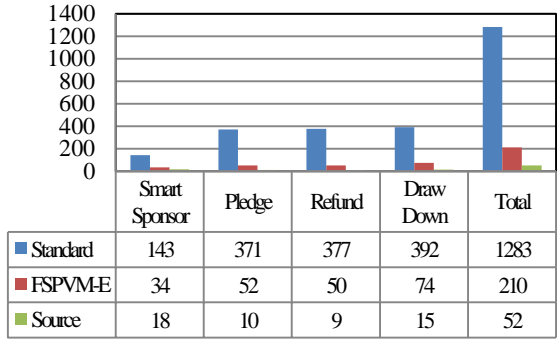

**Verification Workload Comparation for SSCs**

| | Smart Sponsor | Pledge | Refund | Draw Down | Total |
|---|---|---|---|---|---|
| Standard | 143 | 371 | 377 | 392 | 1283 |
| FSPVM-E | 34 | 52 | 50 | 74 | 210 |
| Source | 18 | 10 | 9 | 15 | 52 |

**Fig.24. Comparison of the verification workloads of a conventional theorem proving approach and FSPVM-E for SSCs.**

Fig.24 shows that, compared with our previous work for SSC verification [45], the workload of FSPVM-E has been significantly reduced. Here, the workload of the formalization and verification processes in our previous work are 1283 lines of Coq, which represents a source code size to proof code size ratio of about 27: 1. However, the manual workload in FSPVM-E is only 210 lines of Coq, which represents a source code size to proof code size ratio of about 4: 1.

***High Evaluation Efficiency.*** Although high level automated verification is very important, the computing efficiency of the formal verification engine is also an essential evaluation criterion. Our previous version of FEther obtained good computing efficiency during verification relative to most verification tools for which the computing efficiency is generally quite low. However, we found that *call-by-name termination* (CBNT), *information redundancy explosion* (IRE), and *concurrent reduction* (CR) problems could greatly reduce the computational efficiency of FEther when verifying large programs. Therefore, we have applied the optimization algorithms in [33] that are dedicated to solving these problems to optimize FEther. Briefly, if the CBNT, IRE, and CR problems are triggered simultaneously, the evaluation strategy of higher-order logic proof assistants will cause FEther to employ the entire $Program_{formal}$ rather than a single statement as an evaluation unit, which generates a very large amount of redundant logic information. This condition is summarized by the formulae 26 where $c_{nosub}$ represents the average number of constructors without sub-branches, $c_{sub}$ represents the average number of constructors with sub-branches, $c_{sub_j}$ and $c_{nosub_j}$ represent the number of sub-branches of $c_{sub_i}$ $(i, j \in \mathbb{N}, 0 \le i \le j)$, $r_0$ to $r_i$ represent the number of values constructed by different datatypes, and $size_0$ to $size_i$ represent the number of constructors for each respective datatype. In addition, $infor_{size}$ also contains basic expressions and definitions that can be evaluated directly, and the average number of these basic expressions and definitions are defined as $es_i$ and $ds_i$, respectively.

$$infor_{size} \equiv \left( \sum_{i=0}^{n} \left( es_i + ds_i + c_{nosub_j} + [r_0 \quad \cdots \quad r_i] * \begin{bmatrix} size_0 \\ \cdots \\ size_i \end{bmatrix} \right) * c_{sub_{i-1}}! \right) + size(program_{formal}) \qquad (26)$$

The previous version of FEther was developed strictly following standard interpreter tutorials that explicitly declare sequence statement semantics. As such, the basic functionality of this version can be illustrated by the following simple conditional statement of $obligation_{throw}$ rule 27.

$$obligation_{throw} \stackrel{\text{def}}{=}$$
$$\forall (s, s': statement), if\ (true)\{\ throw\ ();\ \}else\ \{s;\}s'\quad (27)$$

This simple code segment will execute $throw\ ()$ to throw out an executing program and return the initial memory state $m_{init}$ when the condition ($value == 0 \vee completed \vee refnd$) is evaluated as true. However, the evaluation process of $obligation_{throw}$ generates 10,736 lines of logic expressions, and, as shown in Fig.25, executing (i.e., verifying) this very simple code segment using the non-optimized development of FEther requires an execution time of 92.546 s, which is unacceptably long.

In response to this issue, we optimized FEther [32] by applying three proposed optimization schemes, including *redefining semantics*, *deeply embedding*, and *multiple pumps*. This optimized version of FEther is that employed for the proposed FSPVM-E discussed in the present report. The current version of FEther takes a single statement as an evaluation unit, as indicated by the rule 28. Here, we assign $es_\Gamma$ as the number of basic optimal expressions in the context $\Gamma$, and assign $ds_\Gamma$ and $ds_g$ as the number of all bound names of definitions and general definitions in the context $\Gamma$, which are the entry points of the respective definition bodies.

$$infor_{size} \equiv es_\Gamma + ds_\Gamma + ds_g + size(statement)\quad (28)$$

In this manner, the logic information size of an evaluation unit is maintained within a stable range without being affected by the size of the program. A comparison of the result given in Fig.25 for the verification of $obligation_{throw}$ using the non-optimized version of FEther with the result given in Fig.26 using the current version of FEther indicates that the symbolic execution time was decreased from 92.546 s to 0.035 s. As such, the optimized version requires just 3/10000 of the time required by the non-optimized version. In fact, the entire example verifications presented in Subsection V.A required less than 1 s. In addition, we also compared the results obtained with the optimized and non-optimized versions of FEther under equivalent experimental environments with an equivalent data set extracted from a previous study. Under the experimental environment given above, each computer executed the same data set 150 times to obtain the average number of executions required by the FEther evaluation process. In addition, we set an execution time limit of 3600 s to obtain sufficient experimental data. The experimental environments included a specific initial state and an abstract initial state, whose initial arguments of the programs are defined inductively using quantifiers (such as $\forall$ and $\exists$) to logically express all possibilities. As shown in Fig.27, the average execution times of the non-optimized version of FEther with both the specific initial state (SpecNOp) and the abstract initial state (AbsNOp) increase rapidly with an increasing number of program lines, and exceed the time limitation of 3600 s after executing about 30 and 20 program lines, respectively. Moreover, the ranges of fluctuations in the execution times are large, as indicated by the error bars. In contrast, the optimized version of FEther with both the specific initial state (SpecOp) and the abstract initial state (AbsOp) exhibit a steadily increasing average execution time with respect to an increasing number of program lines. In addition, the error bars are much smaller in this case. These results indicate that the execution efficiency of the current version of FEther far exceeded that of the previous non-optimized version of FEther developed in Coq in accordance with the standard tutorial.

**Fig.25.** Execution time of the original non-optimized version of FEther for the simple conditional statement pledge obligation given in (35).

**Fig.26.** Execution time of the optimized version of FEther for the simple conditional statement pledge obligation given in (35).

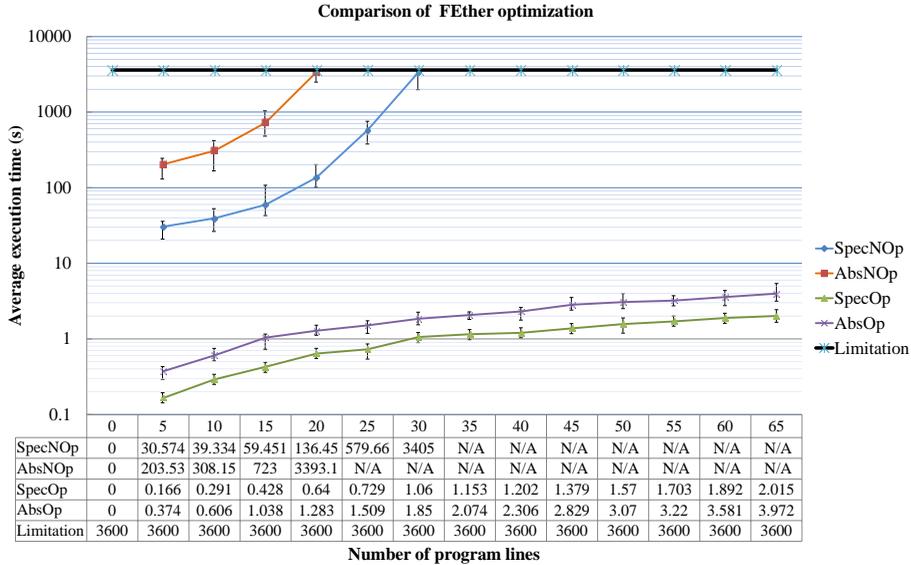

**Fig.27. Comparison of the average evaluation times of optimized (Op) and non-optimized (NOp) FEther under specific initial states (Spec) and abstract initial states (Abs) for previously published example smart contracts [4] as a function of the number of program lines.**

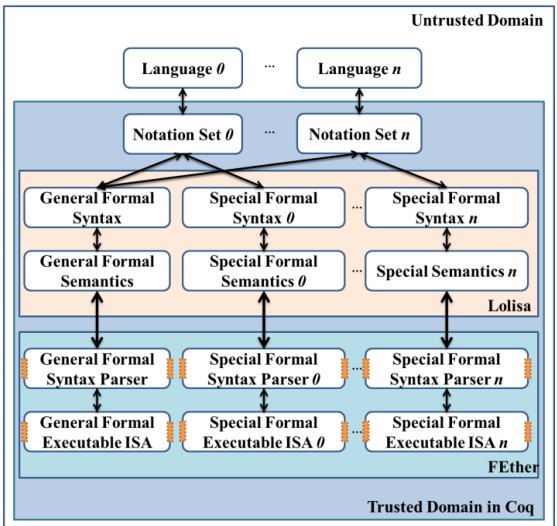

**Fig.28. Architecture for extending the core components of FSPVM-E to support other general-purpose programming languages of multiple blockchain platforms.**

## VI. EXTENSIBILITY AND UNIVERSALITY

While ensuring that the developed Ethereum-based verification platform faithfully captures the intended service properties of smart contracts written in Solidity is essential, further ensuring that this development can be applied to multiple blockchain platforms is also of great value. Therefore, implementing extensibility and universality in the FSPVM-E design was a goal considered from the beginning of its development.

First, we note that the implementations of the GERM framework and the assistant tools and libraries are not dependent on high-level specifications. Therefore, these components of FSPVM-E can be extended by adding new independent types, function specifications, and theorems to support new requirements. The means of facilitating this extensibility were briefly introduced in previous sections. Thus, the extensibility and universality of FSPVM-E are highly dependent on the extensibility of Lolisa and FEther. We deliberately incorporated sufficient extensible space in Lolisa and FEther to extend features, such as pointer formalization and the implementation of independent operator definitions. The architecture of the proposed preliminary scheme for extending Lolisa and FEther to other general-purpose programming languages is illustrated in Fig.28. Extensibility is further realized by the independence of syntax inductive predicates in the same level, which is further supported in the definitions of the semantics. Therefore, Lolisa can be extended to incorporate the features of other languages by adding new typing rule constructors in the formal abstract syntax and the corresponding formal semantics in FEther. As shown in Fig.28, except for the accommodation of specific Solidity data structures, such as contracts and mapping, the remainder of the Lolisa syntax definitions and semantics were designed to be universally applicable to any other general-purpose programming language. Furthermore, Lolisa and FEther were designed based on the GERM framework, which are appropriate for the formalization of any programming language. Finally, as discussed previously, the complex Lolisa formal syntax is hidden in the syntactic abbreviations, which improves user-friendliness significantly.

We treat Lolisa as the core formal language, which is transparent for real-world users, and we logically classify the formal syntax and semantics of Lolisa according to a general component $\mathcal{G}$ and $n$ special components $\mathcal{S}_i$,

TABLE 5
FEATURE COMPARISONS OF FETHER SEMANTICS AND EXISTING SOFTWARE QUALITY TOOLS

| Tool | Spec. | Exec. | Certif. | Verif. | Debug. | Gas | Level | Logic | Hybrid |
|------|-------|-------|---------|--------|--------|-----|-------|-------|--------|
| Lem spec | Yes | Yes | Testing | Yes | No | No | Byte code | Higher order | No |
| Mythril | Yes | Yes | Testing | Yes | Yes | Yes | Byte code | First order | No |
| Hsevm | No | Yes | Testing | No | Yes | No | Byte code | None | No |
| Scilla | Yes | No | Testing | Yes | No | No | Intermediate | Higher order | No |
| Solidity* | Yes | Yes | Verifying | No | No | No | Intermediate | First order | No |
| Cpp-ethereum | No | Yes | Testing | No | No | No | Byte code | None | No |
| KEVM | Yes | Yes | Testing | Yes | Yes | Yes | Byte code | First order | Yes |
| FSPVM-E | Yes | Yes | Verifying | Yes | Yes | Yes | Solidity | Higher order | Yes |

as defined by the rule 29.

$$Lolisa \overset{\text{def}}{=} \mathcal{G} \cup (\cup_{i=0}^{n} \mathcal{S}_i) \tag{29}$$

As a result, a general-purpose programming language $\mathcal{L}_i$ can be formalized by the Lolisa subset $\mathcal{G} \cup \mathcal{S}_i$ by encapsulating the subset using *notation* as a symbolic abbreviation $\mathcal{N}_i$ for $\mathcal{L}_i$, which adopts syntax symbols that are nearly equivalent to the original syntax symbols of $\mathcal{L}_i$. With this method, each $\mathcal{L}_i$ has a corresponding notation set $\mathcal{N}_i$ that satisfies $\mathcal{N}_i \subseteq Lolisa$. This relation is defined by the rule 30.

$$\forall i \in \mathbb{N}. \mathcal{L}_i \leftrightarrow \mathcal{N}_i \equiv \mathcal{G} \cup \mathcal{S}_i \tag{30}$$

As the corresponding definitional interpreter of Lolisa, FEther inherits the extensibility advantages of Lolisa. At the same level, any executable semantic $\mathcal{S}_{exe}$ is independent of any other semantics, and all same-level semantics are encapsulated into an independent module Therefore, FEther is also extendible to new executable semantics to support new abstract syntax of Lolisa without affecting the old semantics.

## VII. COMPARISON WITH RELATED WORK
Compared with most recent tools based on symbolic execution, FSVPM-E supports symbolic execution in higher-order logic systems. Benefiting from this feature, FSPVM-E not only can flexibly formalize and verify properties of Ethereum-based services with Hoare style, but also obtains higher evaluation efficiency and degree of automation compared to standard higher-order logic theorem proving technologies. Moreover, FSPVM-E provides a formal memory model, a virtual symbolic execution environment, and single statement unit evaluation, which solves the *memory*, *environment*, and *constraint solving* problems of conventional tools based on symbolic execution. Moreover, FSPVM-E is first the verification system that provides a large subset of Solidity formal syntax and semantics with GADTs, and it is also the first proof virtual machine can directly execute and verify Solidity programs in Coq.

To compare differences between FSPVM-E and the related works that have pulished open-source tools introduced in Section I, more intuitively, we summarized the differences in Table 5. The compared features are listed and defined as follows:

- Spec.: suitable as a formal specification of the EVM language;
- Exec.: executable on concrete tests;
- Certif.: certifiable self-correctness;
- Verif.: verifiable properties of EVM programs;
- Debug.: provision of an interactive EVM debugger;
- Gas: tools for analyzing the gas complexity of an EVM program;
- Level: analysis or verification level of code;
- Logic: types of essential logic supported;
- Hybrid: support for hybrid verification methods.

## VIII. CONCLUSIONS AND FUTURE WORK
We have presented a hybrid formal verification system denoted as FSPVM-E, which combines symbolic execution and higher-order logic theorem proving for verifying the security and reliability of Ethereum-based smart contract services. An analysis of past studies indicates that the present work represents the first hybrid formal verification system implemented in Coq for Ethereum-based smart contracts that is applied at the Solidity source code level. The foundation of FSPVM-E is a general formal memory model that simulates real world memory operations and provides a basic pointer arithmetic mechanism. The source code of FSPVM-E is Lolisa, which is a large subset of the Solidity programming language. Lolisa is strongly typed according to GADTs, and includes nearly all of the syntax in Solidity. Therefore, the two languages are equivalent, which solves the consistency problem in formalization. The execution engine of FSPVM-E is FEther, which supports static, concolic, and selective symbolic execution simultaneously. The execution engine is driven by the executable formal semantics of Lolisa, and is able to strictly simulate the service behaviors of smart contracts. In addition, we provide two assistant tools and two assistant libraries to increase the degree of automation in the FSPVM-E verification process. Specifically, the two assistant tools are a translator and a generator, which are respectively applied to automatically translate Solidity into Lolisa, and generate specific formal specifications based on a specified execution environment. The two assistant libraries include a static analysis library and an automation tactic library, which are respectively used to scan for standard vulnerabilities in smart contracts, and provide fully-automatic execution (verification) tactics. In addition, we ensured that FSPVM-E is a trusted verification engine by verifying the self-correctness of the core components of FSPVM-E in Coq. We have also provided simple case studies to demonstrate the novel features of FSPVM-E, which include hybrid analysis functions, consistency, automation, and high evaluation efficiency. Finally, the extensibility and universality of FSPVM-E was demonstrated, and a preliminary scheme was proposed to systematically extend FSPVM-E to support the verification of multiple blockchain platforms. As a result, we can now directly verify Ethereum-based service contracts with high automation and evaluation efficiency in Coq directly.

In the future, the Solidity language is still under development and is therefore changing quickly, thus the Lolisa language will be updated quite often in order to cope for that. Besides, we expect to further develop FSPVM-E until it is sufficiently powerful and friendly for use by general programmers to automatically verify the properties of their smart contract programming. First, we are working to introduce support for other Solidity features, such as inline assembly and float datatype. Second, we will develop a trusted translator in Coq directly and certify the correctness of it. Third, we will attempt to implement support for the EOS (https://eos.io/) blockchain platform. Finally, we will design a new verification strategy to improve the automation of modular verification in FSPVM-E.

## APPENDIX

```
contract smartSponsor {
    address public owner;
    address public benefactor;
    bool public refunded;
    bool public complete;
    bool public refnd;
    bool public drwbck;
    uint public numPledges;
    struct Pledge {
        uint amount;
        address eth_address;
        bytes32 message;
    }
    mapping(uint => Pledge) public pledges;
    // constructor
    function smartSponsor(address _benefactor) {
        owner = msg.sender;
        numPledges = 0;
        refunded = false; complete = false;
        refnd = false; drwbck = false;
        benefactor = _benefactor;
    }
    // add a new pledge
    function pledge(bytes32 _message) {
        if (msg.value == 0 || complete || refunded)
            throw;
        pledges[numPledges]
        = Pledge(msg.value, msg.sender, _message);
        numPledges++;
    }

    // refund the backers
    function refund() {
        int i = numPledges − 1;
        if (msg.sender != owner || complete || refunded)
            throw;
        if (drwbck) throw;
        refnd = true;
        while (0 < numPledges) {
            if (pledges[i]. address.send(pledges[i].amount)) {
                numPledges--;
                i--;
            }
            else throw;
        }
        refunded = true;
        complete = true;
    }
    // send funds to the contract benefactor
    function drawdown() {
        if (msg.sender != owner || complete || refunded)
            throw;
        if (refnd) throw;
        drwbck = true;
        if (benefactor.send(this.balance))
            complete = true;
        else throw;
    }
}
```

**Fig.A. 1. The source code of Smart Sponsor contract.**

## Acknowledgment

The authors wish to thank Marisa and Zisheng Wang for their kind assistance and LetPub (www.letpub.com) for its linguistic assistance during the preparation of this manuscript.

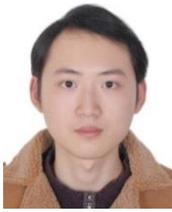

**Zheng Yang** is a Ph.D. student in School of Information and Software Engineering, University of Electronic Science and Technology of China. He received his Bachelor Degree, and became a Ph.D. candidate at School of Information and Software Engineering, University of Electronic Science and Technology of China in 2017. His research interests include programming language theory, formal methods and program verification.

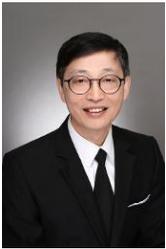

**Hang Lei** received his Ph.D. degree in computer science from the University of Electronic Science and Technology of China, China, in 1997. After graduation, he conducted research in the fields of real-time embedded operating system, operating system security, and program verification, as a professor with the Department of Computer Science, University of Electronic Science and Technology of China. He is currently a professor (doctoral supervisor) in School of Information and Software Engineering, University of Electronic Science and Technology of China. His research interests include big data analytics, machine learning and program verification.

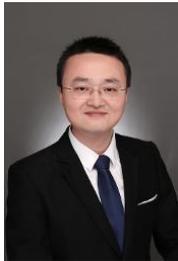

**Weizhong Qian** received his Ph.D. degree in computer science from the University of Electronic Science and Technology of China, China, in 2001. After graduation, he conducted research in the fields of machine learning, software security, and program verification, as an associate professor with the Department of Computer Science, University of Electronic Science and Technology of China. He is currently an assistant professor in School of Information and Software Engineering, University of Electronic Science and Technology of China. His research interests include big data analytics, machine learning and software security.